\documentclass[aps,prl,twocolumn,superscriptaddress,showpacs,preprintnumbers,amsmath,amssymb]{revtex4}
\usepackage{graphicx} 
\usepackage{dcolumn}  

\graphicspath{{ps}}

\newcommand{\ee}{e^{+}e^{-}}
\newcommand{\leplep}{\ell^{+}\ell^{-}}
\newcommand{\jp}{J/\psi}

\newcommand{\psip}{\psi^{\prime}}

\newcommand{\mumu}{\mu^{+}\mu^{-}}
\newcommand{\pipi}{\pi^{+}\pi^{-}}

\newcommand{\bbar}{B\bar{B}}
\newcommand{\dstrbar}{\bar{D}^{*0}}

\newcommand{\Mbc}{M_{\rm bc}}
\newcommand{\DE}{\Delta E}

\newcommand{\rt}{\rightarrow}

\newcommand{\etal}{\em et al.}
\newcommand{\jpsi}{J/\psi}

\begin{document}



\title{ \quad\\[0.5cm] Bounds on the width, mass difference and other properties of $X(3872)\rt \pipi\jp$ decays }

\affiliation{Budker Institute of Nuclear Physics SB RAS and Novosibirsk State University, Novosibirsk 630090}
\affiliation{Faculty of Mathematics and Physics, Charles University, Prague}
\affiliation{University of Cincinnati, Cincinnati, Ohio 45221}
\affiliation{Justus-Liebig-Universit\"at Gie\ss{}en, Gie\ss{}en}
\affiliation{Gifu University, Gifu}
\affiliation{Gyeongsang National University, Chinju}
\affiliation{Hanyang University, Seoul}
\affiliation{University of Hawaii, Honolulu, Hawaii 96822}
\affiliation{High Energy Accelerator Research Organization (KEK), Tsukuba}
\affiliation{Hiroshima Institute of Technology, Hiroshima}
\affiliation{Indian Institute of Technology Guwahati, Guwahati}
\affiliation{Indian Institute of Technology Madras, Madras}
\affiliation{Institute of High Energy Physics, Chinese Academy of Sciences, Beijing}
\affiliation{Institute of High Energy Physics, Vienna}
\affiliation{Institute of High Energy Physics, Protvino}
\affiliation{INFN - Sezione di Torino, Torino}
\affiliation{Institute for Theoretical and Experimental Physics, Moscow}
\affiliation{J. Stefan Institute, Ljubljana}
\affiliation{Kanagawa University, Yokohama}
\affiliation{Institut f\"ur Experimentelle Kernphysik, Karlsruher Institut f\"ur Technologie, Karlsruhe}
\affiliation{Korea Institute of Science and Technology Information, Daejeon}
\affiliation{Korea University, Seoul}
\affiliation{Kyungpook National University, Taegu}
\affiliation{\'Ecole Polytechnique F\'ed\'erale de Lausanne (EPFL), Lausanne}
\affiliation{Faculty of Mathematics and Physics, University of Ljubljana, Ljubljana}
\affiliation{University of Maribor, Maribor}
\affiliation{Max-Planck-Institut f\"ur Physik, M\"unchen}
\affiliation{University of Melbourne, School of Physics, Victoria 3010}
\affiliation{Nagoya University, Nagoya}
\affiliation{Nara Women's University, Nara}
\affiliation{National Central University, Chung-li}
\affiliation{National United University, Miao Li}
\affiliation{Department of Physics, National Taiwan University, Taipei}
\affiliation{H. Niewodniczanski Institute of Nuclear Physics, Krakow}
\affiliation{Nippon Dental University, Niigata}
\affiliation{Niigata University, Niigata}
\affiliation{University of Nova Gorica, Nova Gorica}
\affiliation{Osaka City University, Osaka}
\affiliation{Pacific Northwest National Laboratory, Richland, Washington 99352}
\affiliation{Panjab University, Chandigarh}
\affiliation{Research Center for Nuclear Physics, Osaka}
\affiliation{University of Science and Technology of China, Hefei}
\affiliation{Seoul National University, Seoul}
\affiliation{Sungkyunkwan University, Suwon}
\affiliation{School of Physics, University of Sydney, NSW 2006}
\affiliation{Tata Institute of Fundamental Research, Mumbai}
\affiliation{Excellence Cluster Universe, Technische Universit\"at M\"unchen, Garching}
\affiliation{Toho University, Funabashi}
\affiliation{Tohoku Gakuin University, Tagajo}
\affiliation{Tohoku University, Sendai}
\affiliation{Department of Physics, University of Tokyo, Tokyo}
\affiliation{Tokyo Institute of Technology, Tokyo}
\affiliation{Tokyo Metropolitan University, Tokyo}
\affiliation{Tokyo University of Agriculture and Technology, Tokyo}
\affiliation{CNP, Virginia Polytechnic Institute and State University, Blacksburg, Virginia 24061}
\affiliation{Yonsei University, Seoul}
  \author{S.-K.~Choi}\affiliation{Gyeongsang National University, Chinju} 
  \author{S.~L.~Olsen}\affiliation{Seoul National University, Seoul}     
 \author{K.~Trabelsi}\affiliation{High Energy Accelerator Research Organization (KEK), Tsukuba} 
  \author{I.~Adachi}\affiliation{High Energy Accelerator Research Organization (KEK), Tsukuba} 
  \author{H.~Aihara}\affiliation{Department of Physics, University of Tokyo, Tokyo} 
  \author{K.~Arinstein}\affiliation{Budker Institute of Nuclear Physics SB RAS and Novosibirsk State University, Novosibirsk 630090} 
  \author{D.~M.~Asner}\affiliation{Pacific Northwest National Laboratory, Richland, Washington 99352} 
  \author{T.~Aushev}\affiliation{Institute for Theoretical and Experimental Physics, Moscow} 
  \author{A.~M.~Bakich}\affiliation{School of Physics, University of Sydney, NSW 2006} 
  \author{E.~Barberio}\affiliation{University of Melbourne, School of Physics, Victoria 3010} 
% ...
\collaboration{The Belle Collaboration}



\begin{abstract}
We present results from a study of $X(3872)\rt\pi\pi\jp$ decays
produced via exclusive $B\rt K X(3872)$ decays.
 We determine the mass to be
$M_{X(3872)}= (3871.84\pm 0.27 {\rm (stat)}\pm 0.19~{\rm (syst)})$~MeV,
a 90\% CL upper limit on the natural width of $\Gamma_{X(3872)}<1.2$~MeV,
the product branching fraction
${\mathcal B}(B^+\rt K^+ X(3872))\times {\mathcal B}(X(3872)\rt\pipi\jp) = 
(8.61\pm 0.82~{\rm (stat)} \pm 0.52~{\rm (syst)})\times 10^{-6}$,
and a ratio of branching fractions
${\mathcal B}(B^0\rt K^0 X(3872))/
{\mathcal B}(B^+\rt K^+ X(3872)) = 0.50 \pm 0.14~{\rm (stat)} \pm 0.04~{\rm (syst)}.$
The difference in mass between the $X(3872)\rt\pipi\jp$ signals in $B^+$ and $B^0$ decays
is $\Delta M_{X(3872)}= (-0.69 \pm 0.97 {\rm (stat)} \pm 0.19 {\rm (syst)})$~MeV.
A search for a charged partner of the  $X(3872)$ in the decays $\bar{B}^0\rt K^- X^+$ or $B^+\rt K^0 X^+$,
$X^+\rt\pi^+\pi^0 \jp$ resulted in upper limits on the product branching fractions for these processes
that are well below expectations for the case that the $X(3872)$
is the neutral member of an isospin triplet.
In addition, we examine possible $J^{PC}$ quantum number assignments
for the $X(3872)$ based on comparisons of angular correlations 
between final state particles in $X(3872)\rt \pipi\jp$ decays
with simulated data for $J^{PC}$ values of 
$1^{++}$ and $2^{-+}$.  
We examine the influence of $\rho$-$\omega$ interference in the $M(\pipi)$ spectrum. 
The analysis is based on a  711~fb$^{-1}$ data sample that contains 
772~million $\bbar$ meson pairs collected at the $\Upsilon(4S)$ resonance
in the Belle detector at the KEKB $\ee$ collider.  

\end{abstract}


\maketitle


{\renewcommand{\thefootnote}{\fnsymbol{footnote}}}
\setcounter{footnote}{0}

\section{Introduction}
The $X(3872)$ was first observed by Belle  as a narrow
peak in the $\pipi\jp$ invariant mass distribution
in exclusive  $B^+\rt K^+\pipi\jp$  decays~\cite{skchoi_x3872,conj}.
It was subsequently seen in  $\sqrt{s}=1.96$~TeV
$p\bar{p}$ annihilations 
by CDF~\cite{CDF_x3872} and D0~\cite{D0_x3872} and its
production in $B$ decays was confirmed by BaBar~\cite{babar_x3872}.
A recent summary of the measured properties of the $X(3872)$ is
provided in Tables~10 through~13 of Ref.~\cite{brambilla}.
 
The close proximity of the PDG world-average
of $X(3872)$ mass measurements, 
$M_{\rm avg} = 3871.56 \pm 0.22$~MeV~\cite{PDG},
to the $m_{D^0}+ m_{\dstrbar}$ mass threshold
($3871.8\pm 0.3$~MeV~\cite{PDG})
has engendered speculation that the $X(3872)$
might be a loosely bound $D^0$-$\dstrbar$ {\it molecular}
state~\cite{molecule}.  Theoretical studies of deuteron-like 
$D^0 \dstrbar$ interactions were reported by T\"ornqvist
in 1994, and he predicted bound states for $J^{PC}$ values of
$0^{-+}$ and $1^{++}$~\cite{tornqvist_1994}.  There has been
considerable theoretical interest in the $X(3872)$
line shape in its $D^0\dstrbar$ decay mode~\cite{x3872_lineshape}.
These discussions are constrained by the current uncertainty
in the natural width of the $X(3872)$ in the $\pipi\jp$
decay channel, which is $\Gamma_{X(3872)}<2.3$~MeV (at
the 90\% confidence level)~\cite{skchoi_x3872}.  A
measurement of the natural width in this mode, or an
improvement in the upper limit on its value, would be
useful input to these line-shape studies. 

A close correspondence of the $\pipi$ invariant mass
distribution to expectations for $\rho\rt\pipi$ decays
was reported by Belle~\cite{belle_pipi} and
CDF~\cite{cdf_pipi}. This, together with the observation of
the $X(3872)\rt\gamma\jp$ decay mode by both Belle~\cite{vishnal_qwg7}
and BaBar~\cite{babar_gampsi}, establishes the charge parity 
of the $X(3872)$ as $C=+1$.
A comprehensive study of possible $J^{PC}$
quantum numbers for the $X(3872)$ using a large sample
of $X(3872)\rt\pipi\jp$ decays
was performed by CDF~\cite{CDF_angles,heuser_thesis};
they concluded that only the $1^{++}$ and $2^{-+}$ hypotheses are
consistent with data and other assignments are ruled out
at the $3\sigma$ level or above.  The
$X(3872)\rt\gamma\jp$ decay process would be an allowed
$E1$ transition for a $1^{++}$ assignment and a suppressed
higher multipole for $2^{-+}$; the observation by BaBar and Belle 
of this process favors $1^{++}$~\cite{m2}.  However, a recent BaBar analysis of
the $X(3872)\rt\pipi\pi^0\jp$ decay mode showed some preference for
a $2^{-+}$ assignment~\cite{babar_3pijpsi}.  Since bound molecular 
states are predicted for $J^{PC}=1^{++}$ but not for $2^{-+}$,
an unambiguous experimental determination of the spin-parity
of the $X(3872)$ is an important input to the understanding
of this state.   

Another proposed interpretation for the $X(3872)$ is that it is a
tightly bound diquark-diantiquark four-quark state~\cite{maiani},
in which case two neutral $X(3872)$ states -- orthogonal
mixtures of  $cu\bar{c}\bar{u}$ and $cd\bar{c}\bar{d}$ -- are
expected to exist shifted in mass by $8\pm 3$~MeV.   The
authors of Ref.~\cite{maiani}  suggested that  
these two different  states might result in different
$X(3872)$ masses in the  $B^+\rt K^+\pipi\jp$ and 
$B^0\rt K^0 \pipi\jp$ decay chains.   BaBar measured
the $X(3872)$ properties separately for these two channels and found a mass difference
($\Delta M = 2.7\pm 1.6\pm 0.4$~MeV) that is consistent both with zero and the lower
range of the theoretical prediction~\cite{babar_pipijpsi}.  CDF 
used a comparison of their measured $X(3872)\rt\pipi\jp$ line width
with their experimental resolution to establish
a 95\% CL upper limit of $\Delta M<3.6$~ MeV, for equal production of the
two states~\cite{cdf_pipijpsi}.   These results are not definitive tests
of the prediction of Ref.~\cite{maiani}; the statistical significance of the
BaBar signal for $B^0\rt K^0 X(3872)$ is marginal ($9.4\pm 5.2$ events)
and the interpretation of the CDF limit depends upon the unknown relative production
strengths for the two different states.  Thus, a more precise comparison
of the $X(3872)$ produced in $B^+$ and $B^0$ decays is needed. 

In the  diquark-diantiquark scheme, the $X(3872)$ is expected to be
the $I_3 = 0$ member of an isospin triplet.  Since the dominant 
weak interaction process responsible for $B\rt KX(3872)$ decays
is the isospin conserving $b\rt c\bar{c} s$ transition,
the charged $I_3=\pm 1$ partner states (that decay via 
$X^+\rt \rho^+\jp$) are expected to be produced  in $B$ decays
at a rate that is twice that for the neutral $X(3872)$~\cite{braaten}.  
The BaBar group studied the process $B \rt K \pi^+\pi^0 \jpsi$ and
placed upper limits on the product branching  fractions for 
$X^+\rt\pi^+\pi^0\jp$ that are below isospin
expectations~\cite{babar_pipi0jpsi}.

Here we report on a study of $X(3872)\rt \pipi\jp$ decays 
produced via the exclusive decay $B\rt KX(3872)$. 
We use a 711~fb$^{-1}$ data sample that contains 772~million 
$\bbar$ pairs  collected in the Belle detector at the KEKB
energy-asymmetric  $\ee$ collider~\cite{full_data}.  The data
were accumulated at a
center-of-mass system (cms) energy of $\sqrt{s} = 10.58$~GeV,
at the peak of the $\Upsilon(4S)$ resonance.  
KEKB is described in detail in Ref.~\cite{KEKB}.

\section{Detector description}
The Belle detector is a large-solid-angle magnetic 
spectrometer  that consists of a silicon vertex 
detector (SVD), a 50-layer cylindrical drift chamber (CDC), an 
array of aerogel threshold Cherenkov counters (ACC),  a 
barrel-like arrangement of time-of-flight  scintillation 
counters (TOF), and an electromagnetic calorimeter
(ECL) comprised of CsI(Tl) crystals  located inside
a superconducting solenoid coil that provides a 1.5~T
magnetic field.  An iron flux-return located outside of 
the coil is instrumented to detect $K_L$ mesons and to 
identify muons (KLM).  The detector is described in detail 
elsewhere~\cite{Belle}.

\section{$B\rt K\pipi\jp$ event selection}
We select events that contain a $\jp\rt\leplep$
($\leplep=\ee$ or $\mumu$), either a charged or neutral kaon, 
and a $\pipi$ pair using criteria described in 
Refs~\cite{skchoi_x3872} and~\cite{skchoi_y3940}.  
The leptons from the $\jp\rt\leplep$ decay are required
to pass minimal lepton identification criteria and the
invariant mass of the pair is required to be in the 
ranges $-21$~MeV~$\le(M_{\mumu}-m_{\jp})\le 20$~MeV and
$-24$~MeV~$\le (M_{\ee}-m_{\jp})\le 20$~MeV,
where $m_{\jp} = 3096.92\pm 0.01$~MeV
is the world-average value for the $\jp$ mass~\cite{PDG}.
For $\jp\rt \ee$ candidates, photons within 50~mrad of the
$e^+$ and/or $e^-$ tracks are included in the invariant mass
calculation.  The number of events with multiple $\jp$ candidates
is negligibly small.
Candidate $K^+$ mesons are
charged tracks with a kaon identification likelihood that is 
higher than that for a pion or a proton;
neutral kaons are detected in the $K_S\rt\pipi$ decay channel using
the $K_S$ selection criteria described in Ref.~\cite{belle_ffang}.
The charged pions are required to have a pion likelihood greater
than that of a kaon or a proton.  
Some events have more than one acceptable combination of hadron tracks. 
In these cases, which include 3\% of the events in the signal region,
the tracks with the best vertex fits are used. 
To reduce the level of $\ee\rt q\bar{q}$ ($q=u,d,s~{\rm or}~c$-quark)
continuum events in the sample,
we also require  $R_2 < 0.4$, where $R_2$ is the normalized
Fox-Wolfram moment~\cite{fox}.

Events that originate from $B \rt K\pipi\jp$ decays are identified by the 
cms energy difference  $\DE\equiv E_B^{\rm cms} - E_{\rm beam}^{\rm cms}$
and the beam-energy-constrained mass 
$\Mbc\equiv\sqrt{(E_{\rm beam}^{\rm cms})^2-(p_B^{\rm cms})^2}$,
where $E_{\rm beam}^{\rm cms}$ is the cms beam
energy, and $E_B^{\rm cms}$ and $p_B^{\rm cms}$ are the cms 
energy and momentum of the $K\pipi\jp$ combination.  We select 
events with $\Mbc>5.20$~GeV and $-0.15~{\rm GeV}<\DE<0.2$~GeV.
We define
signal regions as  $5.272~{\rm GeV} < \Mbc <5.286$~GeV and
$-0.35\,\mathrm{GeV} \le \DE \le 0.03\,\mathrm{GeV}$;
these correspond to $\simeq \pm 2.5\sigma$ windows around
the central values for each variable.

In addition to selecting $B\rt K X(3872)$ events,
these selection criteria isolate a rather pure sample of
$B\rt K\psip$, $\psip\rt\pipi\jp$ events~\cite{psiprime}.
These events are used as a calibration reaction to 
determine the $\Mbc$, $\DE$ and $M(\pipi\jp)$ peak 
positions and resolution values, and to validate the
Monte Carlo-determined acceptance calculations.

For each event we compute $M(\pipi\jp)$ from the relation
\begin{equation}
M(\pipi\jp)=M^{\rm meas}_{\pipi\leplep}
-M^{\rm meas}_{\leplep}+ m_{\jp},
\label{mpipijp_def}
\end{equation}
where $M^{\rm meas}_{\pipi\leplep}$ and $M^{\rm meas}_{\leplep}$
are the measured $\pipi\leplep$ and $\leplep$ invariant masses,
respectively.   For studies of the $\psip\rt\pipi\jp$ control
sample we use events in the interval
$3.635~{\rm GeV}\le M(\pipi\jp)\le 3.735$~GeV; for $X(3872)$  
studies we use $3.77~{\rm GeV}\le M(\pipi\jp)\le 3.97$~GeV. 
The $M(\pipi\jp)$ signal regions are defined as
$|M(\pipi\jp)-M_{\rm peak}|\le 0.009$~GeV, where $M_{\rm peak}=3.686$~GeV
and $3.872$~GeV for the $\psip$ and $X(3872)$, respectively.
We select events with a dipion invariant mass requirement
of $M_{\pipi} > (M(\pipi\jp) - (m_{\jp} + 150~{\rm MeV}))$,
which corresponds to $M_{\pipi}>625$~MeV for the $X(3872)$
and $>439$~MeV for the $\psip$ events.  After this requirement,
which results in a 6\% signal loss, the background under the
$X(3872)\rt\pipi\jp$ signal peak is relatively flat 
and similar in shape to that under the $\psip\rt\pipi\jp$
peak.

\section{Monte Carlo results}

We use Monte Carlo (MC) simulated events to determine acceptance and to
evaluate possible differences in mass biases for the $\psip$ and $X(3872)$
mass regions~\cite{geant}.
The $\psip$ MC simulation uses an input mass and width of: $m_{\psip}=3686.09$~MeV and
$\Gamma_{\psip}=0.3$~MeV~\cite{PDG}.  The default $X(3872)$
simulation assumes $J^{PC}=1^{++}$ and a $\pipi\jp$ final state that is
entirely $B\rt \rho\jp$ with the $\rho$ and $\jp$ in a relative
$S$-wave~\cite{evtgen}.  The $X(3872)$ mesons are generated with a mass of 
$M^{\rm gen}_{X(3872)}=3871.40$~MeV and zero natural width.
The simulated events are processed through the same reconstruction
and selection codes that are used for the real data.

We perform an unbinned three-dimensional likelihood fit
($\Mbc$ {\it vs.} $M(\pipi\jp)$ {\it vs.} $\DE$)
to the selected data using a single Gaussian function 
for the $\Mbc$ signal probability density function (PDF)  
and an ARGUS function~\cite{ARGUS} as the PDF for the
combinatorial background ({\it i.e.}, backgrounds
where one or more of the tracks used to reconstruct the $B$
originates from the accompanying $\bar{B}$).
For $\DE$ we use a bifurcated Gaussian for
the signal PDF and a second-order polynomial
for the $\DE$ combinatorial background.  For the $M(\pipi\jp)$
signal PDF we use a Breit-Wigner function (BW)  
convolved with a resolution function that is the sum of
a core and tail Gaussian; for the combinatorial background PDF 
we use a third-order polynomial.  For $\psip$ fits in both data
and MC, we fix the BW width at $0.3$~MeV.  For the $X(3872)$
MC fits, we fix the BW width at zero.

In addition to combinatorial background, these criteria select
events of the type $B\rt K_X \jp$, where $K_X$ designates
strange meson systems that decay to $K\pipi$ final states
such as the $K_1(1270)$, $K_2^*(1430)$, etc.~\cite{hulya}.  The
$\Mbc$ and $\DE$ distributions for these events are
the same as those of the $X(3872)$ signal, but they produce a slowly
varying $M(\pipi\jp)$ distribution in the $\psip$ and
$X(3872)$ signal regions.  The $\Mbc$ and $\DE$
PDFs that are used to represent this {\it peaking} background are the same as
those used for the signal and a linear form is used
for its $M(\pipi\jp)$ PDF.

The results of fits to MC samples of $B^+\rt K^+\psip$, $B^0\rt K_S\psip$,
$B^+\rt K^+X(3872)$ and $B^0\rt K_S X(3872)$ are summarized in Table~\ref{tbl:mc}.
In order to facilitate comparisons of the resolution for different decay
channels, the relative fraction of the tail and core Gaussian for all modes 
is fixed at the value returned from the fit to the $K^+\psip$ MC sample (17.7\%). 
This restriction is found to induce negligible differences from the shapes
of the resolution functions that are individually optimized for the other samples.  
While the core resolution width is nearly the same for all channels,
the tail resolution widths for $X(3872)$ decays are significantly higher than those
for the $\psip$, but in both cases the tail widths for the $K^+$ and $K_S$ modes 
are consistent with being the same.
The MC indicates that there are biases in the $M(\pipi\jp)$ measurement
that are smaller for the $X(3872)$ modes than for the $\psip$ modes. 
These are due to a bias in the measurement of the low momentum charged pions.  
The pions from $X(3872)$ decays have, on average, higher momentum than 
those from $\psip$ decays and the $X(3872)$ mass measurement bias is smaller.  In both cases,
the mass measurement biases for the $K^+$ and $K_S$ modes are consistent with being the same.
The results of fits to the combined $K^+$ and $K_S$ modes are also shown in Table~\ref{tbl:mc}.
(The listed efficiencies do not include the $\psip\rt\pipi\jp$,
$\jp\rt\leplep$ or $K_S\rt\pipi$ branching fractions).   

\begin{table}[htb]
\begin{center}
\caption{\label{tbl:mc}
Results from fits to the selected MC event samples.  Here 
$\epsilon=N_{\rm sig}/N_{\rm gen}$ is the detection efficiency, $\sigma_{\rm core}$ and $\sigma_{\rm tail}$ 
are the widths of the core and tail components of the mass resolution
and $M_{\rm gen}-M_{\rm fit}$ are the MC mass measurement biases.  All errors are statistical.}
\begin{tabular}{l|c|c|c|c}\hline
Channel                  & $\epsilon$   & $\sigma_{\rm core}$   & $\sigma_{\rm tail}$  &  $M_{\rm gen}-M_{\rm fit}$ \\
                         &   (percent)  &      (MeV)            &      (MeV)           &        (MeV)               \\\hline\hline
$K^+\psip$               & $17.8 \pm0.2$&   $1.83\pm 0.02$      &  $5.66\pm 0.14$      &  $0.74\pm 0.02$            \\
$K_S\psip$               & $14.1 \pm0.2$&   $1.83\pm 0.03$      &  $6.10\pm 0.21$      &  $0.74\pm 0.03$            \\
Combined                 &              &   $1.84\pm 0.02$      &  $5.66\pm 0.13$      &  $0.72\pm 0.02$            \\\hline
$K^+X(3872)$             & $19.1\pm0.2$ &   $1.93\pm 0.04$      &  $7.69\pm 0.17$      &  $0.60\pm 0.02$            \\
$K_SX(3872)$             & $15.2\pm0.2$ &   $1.89\pm 0.02$      &  $7.64\pm 0.21$      &  $0.64\pm 0.02$            \\
Combined                 &              &   $1.93\pm 0.02$      &  $7.70\pm 0.15$      &  $0.60\pm 0.02$            \\
\hline\hline
\end{tabular}
\end{center}
\end{table}

\section{Fits to the $\psip\rt\pipi\jp$ data samples}

\begin{figure*}[htb]
\mbox{
  \includegraphics[height=0.4\textwidth,width=0.8\textwidth]{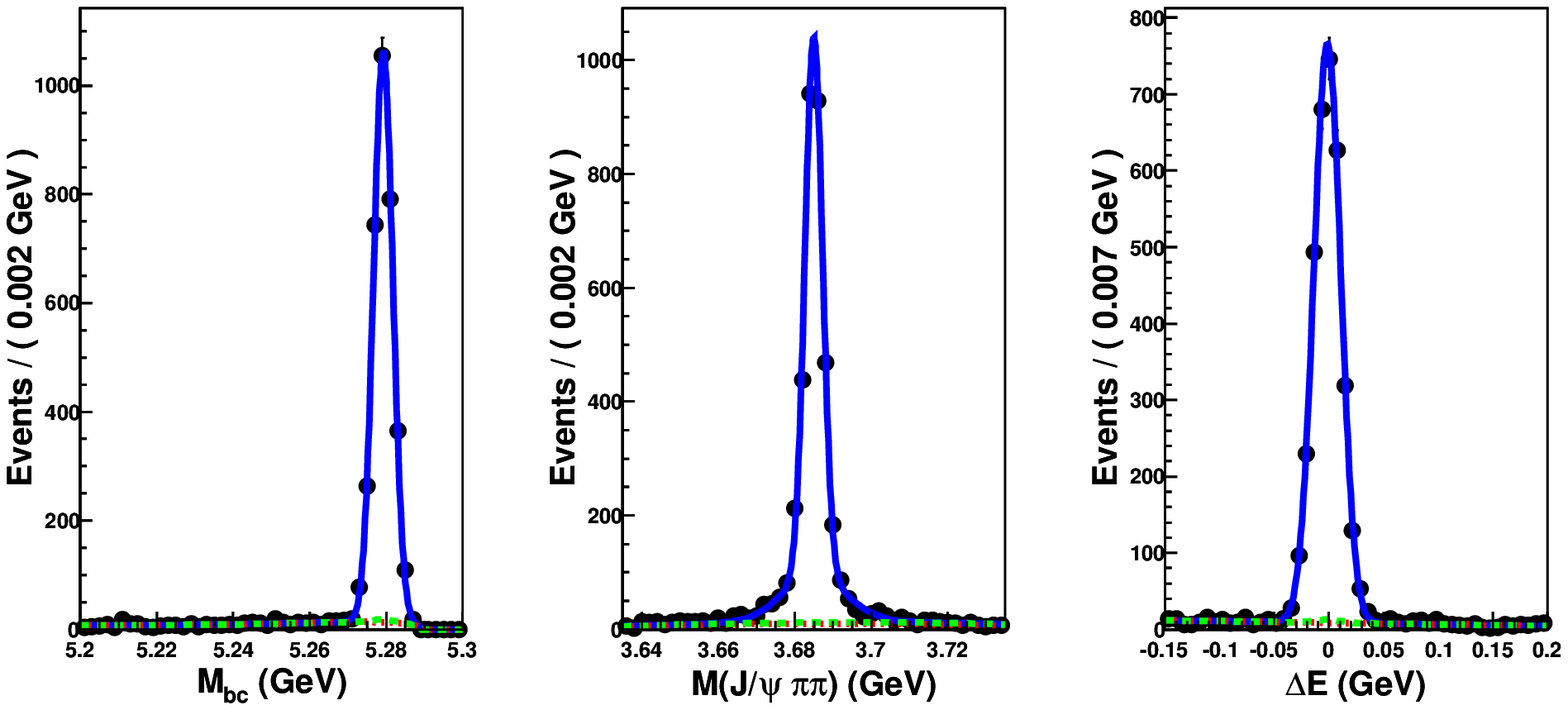}
}
\mbox{
  \includegraphics[height=0.4\textwidth,width=0.8\textwidth]{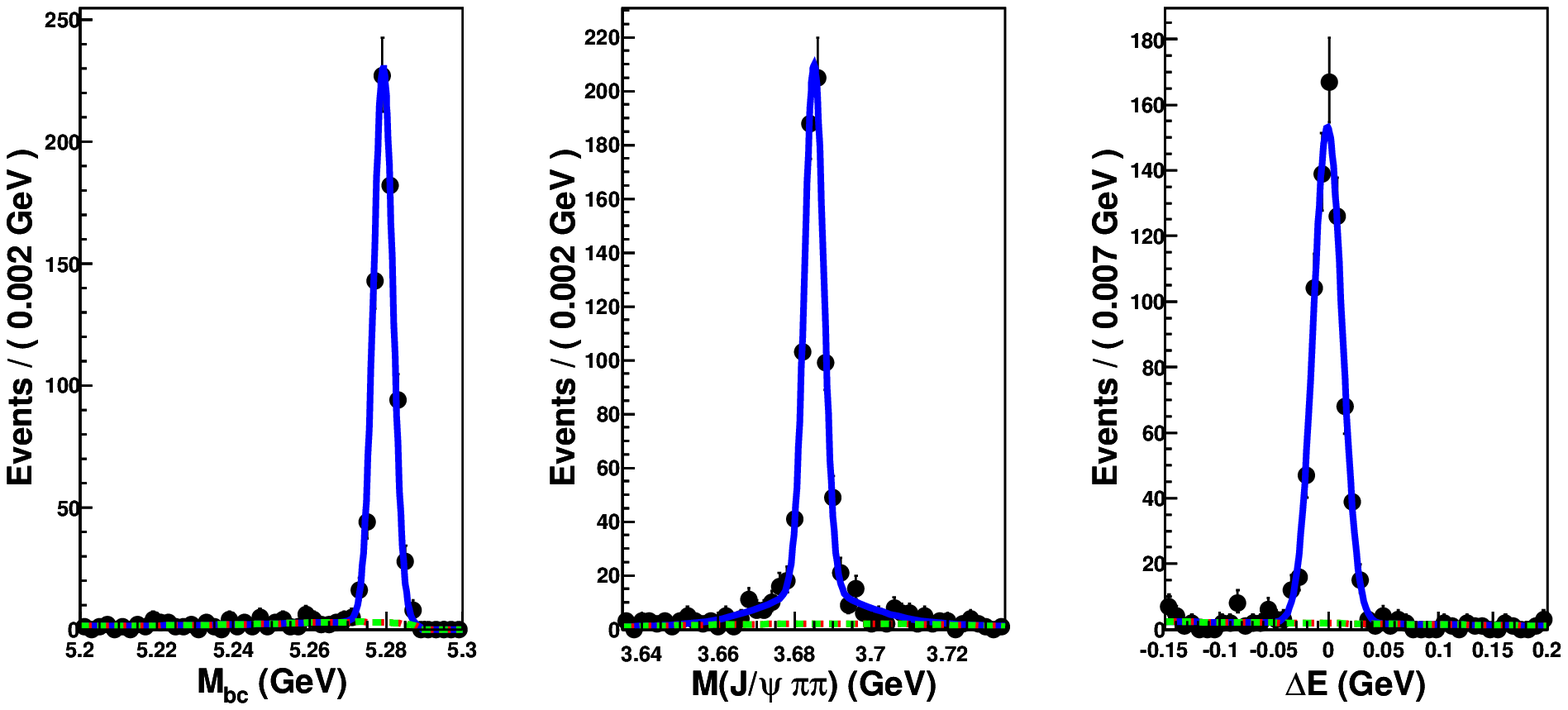}
}
\caption{ 
The $\Mbc$ (left), $M(\pipi\jp)$ (center) and
$\DE$ (right) distributions for  
$B^+\rt K^+\psip$ (top) and $B^0\rt K_S\psip$ (bottom) event candidates within the signal 
regions of the other two quantities.   The curves show
the results of the fits described in the text.
}
\label{fig:mpipijpsi_kpsip_3box}
\end{figure*}

For fits to the $\psip$ data we fix the BW width at $0.3$~MeV and allow
the core and tail widths of the $M(\pipi\jp)$ resolution function to vary
as free parameters.  The results of the fits  to $B^+\rt K^+\psip$ 
($B^0\rt K_S\psip$)  are the  smooth curves in the upper (lower) panels
of Fig.~\ref{fig:mpipijpsi_kpsip_3box}, where $\Mbc$, $M(\pipi\jp)$ and 
$\DE$ distributions for events within the signal regions of the other two
quantities are shown.  In each panel, the combinatorial background is shown
as a (red) dotted line, the combinatorial plus peaking background is shown
as a (green) dashed line and the total background plus signal is shown as
a (blue) solid line. The fit results are summarized in Table~\ref{tbl:psip}.
They show a mass bias, {\it i.e.}, a difference between the fitted mass and the
PDG world-average value for $m_{\psip}$, that is larger than the MC mass
bias, indicating that the MC simulation of the bias in the pion momentum
measurement is imperfect.

\begin{table}[htb]
\begin{center}\caption{\label{tbl:psip}
Results from fits to the $\psip$ event candidates.  Here 
$N^{\rm evts}$ denotes the number of signal events returned from the fit, $\sigma_{\rm core}$
and $\sigma_{\rm tail}$  are the mass resolution parameters, and
$\Delta M_{\rm PDG}= M_{\rm PDG}-M_{\rm fit}$ denotes the mass measurement biases.  
All errors are statistical.}
\begin{tabular}{l|c|c|c|c}\hline
Channel     &$N^{\rm evts}$ &$\sigma_{\rm core}$ & $\sigma_{\rm tail}$ & $\Delta M_{\rm PDG}$ \\
            &               &   (MeV)            &   (MeV)             &  (MeV)               \\
\hline\hline
$K^+\psip$  &$3575\pm 64$   &   $2.25\pm 0.05$   & $ 8.4\pm 0.5$       &   $1.12\pm 0.05$ \\
$K_S\psip$  &$ 814\pm 30$   &   $2.45\pm 0.11$   & $13.8 \pm 1.6$      &   $1.05\pm 0.12$ \\
Combined    &$4367\pm 72$   &   $2.28\pm 0.04$   & $ 8.7\pm 0.5$       &   $1.11\pm 0.05$ \\
\hline\hline
\end{tabular}
\end{center}
\end{table}

As a test of the validity of the MC acceptance calculations, we
determine branching fractions for $B\rt K^+\psip$ and $K_S \psip$
via the relation
\begin{equation}
{\mathcal B }(B\rt K\psip ) =
\frac{ N^{\rm evts}_K} {N_{B\bar{B}}
{\epsilon_{K}  f_{K}
{\mathcal B}_{\psip\rt\pipi\jp}
\mathcal B}_{\jp\rt\ell\ell}},
\label{eq:branch_psip}
\end{equation}
\noindent
where $N^{\rm evts}_K$ is the number of
signal events for $K=K^+$ and $K=K^0$,
$N_{B\bar{B}}=(772\pm 11)\times 10^6$ is the number of 
$B\bar{B}$ events in the data sample, 
${\mathcal B}_{\psip\rt\pipi\jp}=0.336\pm0.004$ and
${\mathcal B}_{\jp\rt\ell\ell}= 0.119\pm 0.001$
(sum of the $\ee$ and $\mumu$ modes) are PDG world-average branching
fractions~\cite{PDG}, $\epsilon_K$ is the efficiency
for the corresponding $K$ channel, $f_{K^+}=1$ and
$f_{K_S}=0.346$~\cite{fks}.
The results are:
  ${\mathcal B}(B^+\rt K^+\psip) = (6.51 \pm 0.12) \times 10^{-4}$ and
  ${\mathcal B}(B^0\rt K^0\psip) = (5.22 \pm 0.19) \times 10^{-4}$,
where only statistical
errors are shown.  The $B^+$ branching fraction result
agrees well with the PDG world-average value of 
$(6.46\pm 0.33)\times 10^{-4}$.
The $B^0$ result is somewhat lower than the PDG value of
($6.2\pm 0.5)\times 10^{-4}$~\cite{PDG}, however, the errors
quoted on the measurements reported here do not include
systematic uncertainties. 

\section{$X(3872)\rt\pipi\jp$ mass, width and product branching fractions}
\begin{figure*}[htb]
\mbox{
  \includegraphics[height=0.4\textwidth,width=0.8\textwidth]{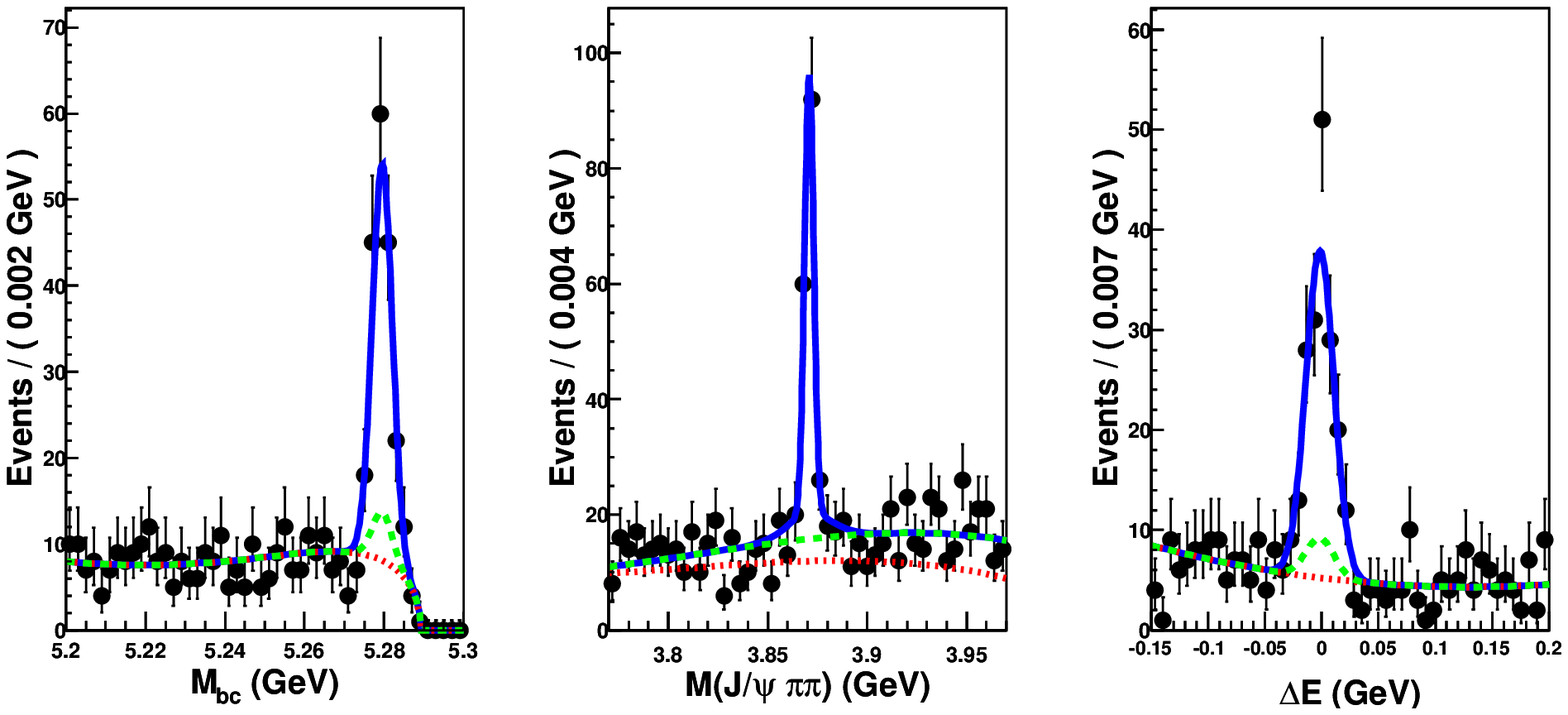}
}
\mbox{
  \includegraphics[height=0.4\textwidth,width=0.8\textwidth]{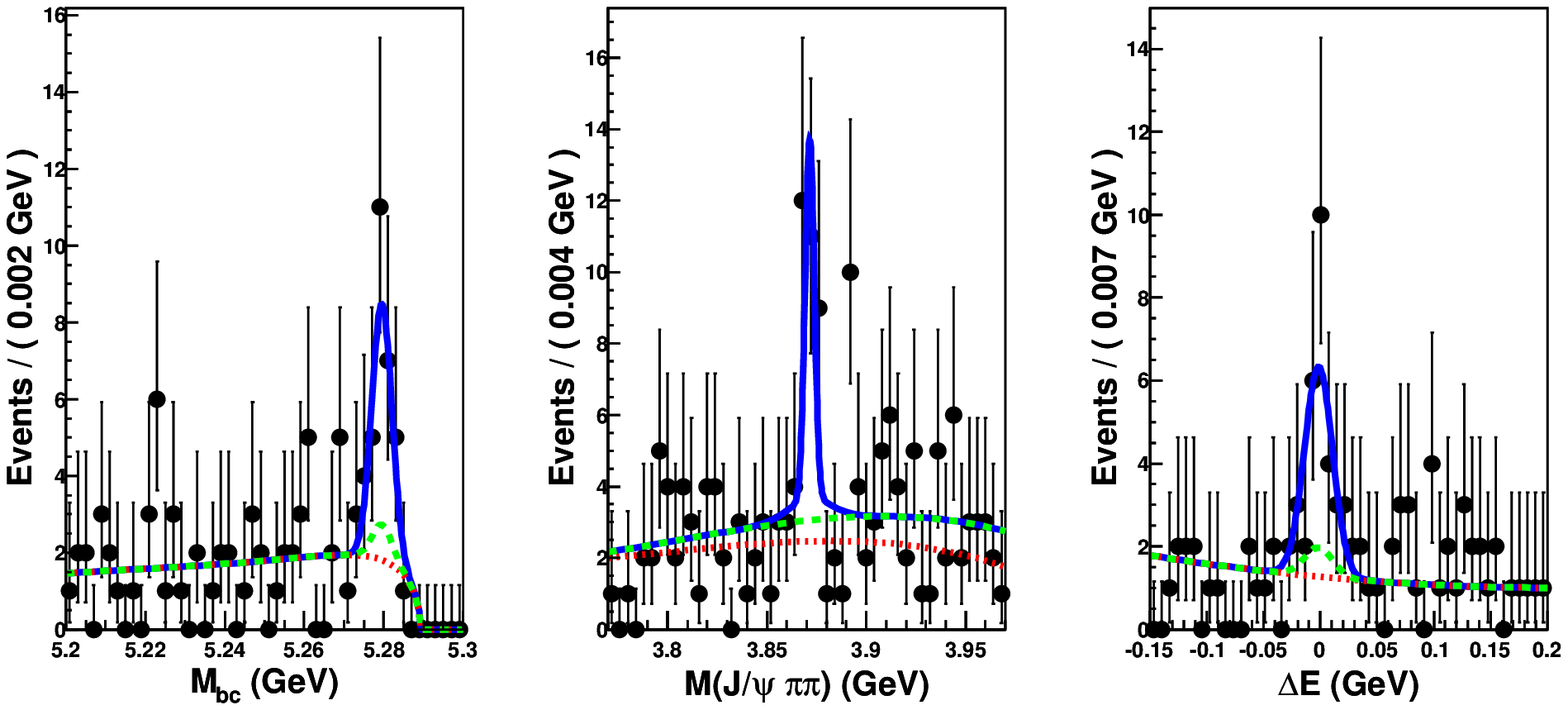}
}
\caption{ 
The $\Mbc$ (left), $M(\pipi\jp)$ (center) and
$\DE$ (right) distributions for  
$B^+\rt K^+ X(3872)$ (top) and $B^0\rt K_S X(3872)$ (bottom) event candidates within the signal 
regions of the other two quantities.   The curves show
the results of the fit described in the text .
}
\label{fig:mpipijpsi_kx3872_3box}
\end{figure*}

The upper panels in Fig.~\ref{fig:mpipijpsi_kx3872_3box} 
show the $\Mbc$, $M(\pipi\jp)$ and $\DE$ distributions for events
within the signal regions of the other two quantities
for the $B^+\rt K^+ X(3872)$ event candidates together with the results
of the fit.   In these fits, the peak mass and full width of the
BW function that represents the $M(\pipi\jp)$ signal are free parameters,
the width of the core  Gaussian resolution function is fixed at
$\sigma_{\rm core}=2.28$~MeV, the measured $\psip$ core resolution value, 
and the width of the tail Gaussian is fixed at $\sigma_{tail}= 11.5$~MeV, 
the tail width from the $\psip$ data sample fit multiplied
by the ratio of the MC-determined $X(3872)$ and $\psip$ tail widths
to account for its $M(\pipi\jp)$ dependence.
The value for $\Gamma_{X(3872)}$ returned from the fit is at its
lowest allowed value of $0.1$~MeV~\cite{lower_limit}.  Other results from the
fit are summarized in Table~\ref{tbl:x3872}.

The lower panels of Fig.~\ref{fig:mpipijpsi_kx3872_3box} show 
the $\Mbc$, $M(\pipi\jp)$ and $\DE$ distributions for events in the signal
regions of the other two quantities for 
the $K_S$ event sample, where an $X(3872)\rt\pipi\jp$ signal is evident.
The results of a fit that fixes the natural width at zero and the resolution widths
at the same values used for the fit to the $K^+X(3872)$ channel
but with the peak mass allowed to vary, are shown as curves in the
figure and summarized in Table~\ref{tbl:x3872}.
The statistical significance of the $X(3872)$ signal yield for the $K_S$ event sample is
6.1$\sigma$.  This is determined from $-2 \ln({\mathcal L}_0 /{\mathcal L}_{\rm max})$, 
where ${\mathcal L_{\rm max}}$ is the maximum likelihood and ${\mathcal L_0}$ is the likelihood
for zero signal yield with the change in the number of degrees of freedom  taken into account.
The difference in mass for the $X(3872)$ state produced in $B^+$ minus that from $B^0$ decays
({\it i.e.,} $\Delta M = M_{+}-M_{0}$)  is 
\begin{equation}
\Delta M_{X(3872)} = (-0.69 \pm 0.97~{\rm (stat)} \pm 0.19~{\rm (syst)})~{\rm MeV}.
\end{equation}  
Although many sources of systematic error on the mass measurement cancel in the this difference,
assumptions on the natural width used in the fit and possible differences in momentum measurement
biases between charged and neutral kaons do not cancel.   We estimate the error associated with the
natural width to be $0.14$~MeV from the change in $\Delta M_{X(3872)}$ determined from a fit to the $K_S$ event
sample that uses a natural width fixed at 3~MeV.  The difference of the measured $\psip$ masses
in the $B^+\rt K^+\psip$ and $B^0\rt  K_S\psip$ channels is $\Delta M_{\psip} = (-0.07\pm 0.13)$~MeV.   
We use the error on $\Delta M_{\psip}$ as an estimate of the systematic error associated with possible
different charged and neutral kaon measurement biases. 

This result strongly disfavors the prediction of Ref.~\cite{maiani}.
The BaBar  measurement for this quantity is ($2.7\pm 1.6\pm 0.4$)~MeV~\cite{babar_pipijpsi}.

\begin{table}[htb]
\begin{center}
\caption{\label{tbl:x3872}
Results from fits to the $X(3872)$ event candidates.  Here $N^{\rm evts}$ are the numbers of
signal events returned from the fit and
$M_{\rm fit}$ is the fitted mass value.  All errors are statistical.}
\begin{tabular}{l|c|c}\hline
Channel          &$N^{\rm evts}$     & $M_{\rm fit}$~(MeV)   \\
\hline\hline
$K^+ X(3872)$     &$151  \pm 15 $  & $3870.85\pm 0.28$   \\
$K^0 X(3872)$     &$ 21.0\pm 5.7$  & $3871.54\pm 0.93$   \\
Combined           &$173  \pm 16 $  & $3870.92\pm 0.27$   \\
\hline\hline
\end{tabular}
\end{center}
\end{table}

\subsection{$M_{X(3872)}$  determination}

Since the mass difference is consistent with zero and the resolution
functions for the $K^+ X(3872)$ and $K^0 X(3872)$ are consistent with
being the same, we
determine an $X(3872)$ mass value from the single fit to the combined 
samples.   To account for the mass measurement bias,
we correct the fitted mass given in Table~\ref{tbl:x3872}
by adding a correction $\delta M = (0.92\pm 0.06$)~MeV,
which is the MC-determined $X(3872)$ mass measurement bias scaled
by the ratio of the measured and MC-determined $\psip$ mass biases.
The validity of this procedure is tested with MC event samples
of narrow resonances with $\psip$ and $X(3872)$ ($J^{PC}=1^{++}$)
decay dynamics at different mass values ranging from $m_{\psip}$
to 3872~MeV.  It is found for both dynamics that the MC mass bias falls
linearly with increasing $M(\pipi\jp)$ with slopes ($b^{MC}$) that
are very nearly equal: $b^{MC}_{\psi}=-0.096\pm 0.004$~keV/MeV and
$b^{MC}_{X(3872)}=-0.097\pm 0.004$~keV/MeV, indicating that 
using the $\psip$ measurement performed
at a mass that is 186~MeV below $M_{X(3872)}$ to scale the 
mass shift near 3872~MeV is reasonable. 

The offset between the
MC-determined $\psip$-like
and $X(3872)$-like mass biases is ($0.053\pm 0.005$)~MeV.  We use this
offset, scaled by the $\psip$ data-MC mass bias ratio, 
as the systematic error associated with the decay model.  The
systematic error associated with the MC modeling of the
low energy pion momentum measurements is determined by 
comparing results from different versions of the MC 
simulation to be $0.15$~MeV.

The result is
\begin{equation}
M_{X(3872)} = (3871.84 \pm 0.27~{\rm (stat)} \pm 0.19~{\rm (syst)})~{\rm MeV},
\end{equation} 
where the systematic error is dominated by the error on the mass bias 
correction  (0.16~MeV)
and uncertainties in the decay dynamics used to generate the MC samples
used to study the mass bias (0.09~MeV). It also includes
the uncertainties in the $\jp$ and $\psip$
masses and the choice of parameterization used in the three dimensional fit. 
The latter is estimated from the quadratic sum of the changes induced by
$\pm 1\sigma $ variations of the fit parameters and from the use of different
functional forms for the PDFs.   The systematic error evaluation
is summarized in Table~\ref{tbl:mass_syst}.

\begin{table}[htb]
\begin{center}
\caption{\label{tbl:mass_syst}
Systematic errors on the mass measurement.}
\begin{tabular}{l|c}\hline
Source               &   Systematic error~(MeV) \\
\hline\hline
$m_{\jp}$             &           0.01           \\
$m_{\psip}$            &           0.04           \\
Bias correction      &           0.16           \\
3-dim. fit model     &           0.03           \\
MC model dependence  &           0.09           \\
\hline
Quadrature sum       &           0.19           \\
\hline\hline
\end{tabular}
\end{center}
\end{table}

\subsection{ $\Gamma_{X(3872)}$ upper limit}

\begin{figure}[htb]
\mbox{
  \includegraphics[height=0.25\textwidth,width=0.5\textwidth]{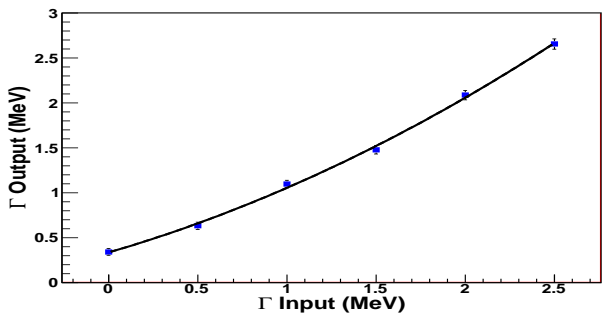}
}
\caption{ Fitted vaules for $\Gamma_{X(3872)}$ (vertical) {\it versus} the MC generator 
input values (horizontal).  The curve is the result of a fit to a second-order
polynomial. }
\label{fig:width_bias}
\end{figure}

The current best limit on the width of the $X(3872)$ is the
90\% confidence level (CL) upper limit of $\Gamma_{X(3872)}<2.3$~MeV
reported in the original discovery paper~\cite{skchoi_x3872}. 
This is narrower than the $M(\pipi\jp)$ mass resolution of 
the Belle detector in the mass region of the $X(3872)$,
$\langle \sigma \rangle \simeq 4$~MeV.   However, the three dimensional
fits used in the analyses reported here are sensitive to
natural widths that are narrower than the resolution because
of the constraints on the area of the $M(\pipi\jp)$ signal 
peak provided by the $\Mbc$ and $\DE$ components.  Because of these 
constraints on the area of the peak, the measured peak height is
sensitive to $\Gamma_{X(3872)}$.   This is
demonstrated in  Fig.~\ref{fig:width_bias}, which shows the
results of fits to high statistics MC samples where the $X(3872)$
is generated with widths ranging from zero to $2.5$~MeV.
Although the measurements have some bias, especially at
very small widths, the different input widths are clearly
distinguishable.   The curve in Fig.~\ref{fig:width_bias}
shows the results of fit of a parabola to the MC measurements.

A fit to the $X(3872)\rt\pipi\jp$ mass peak in data with
$\Gamma_{X(3872)}$ as a free parameter returns a value that is at the 
lower limit imposed on the fit.  To establish an upper limit on its value,
we made a  study of how the fit likelihood depends on $\Gamma_{X(3872)}$. 

\begin{figure}[htb]
\mbox{
  \includegraphics[height=0.22\textwidth,width=0.44\textwidth]{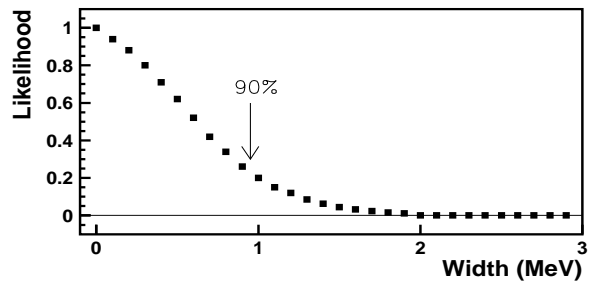}
}
\caption{ 
Likelihood values from the $\Gamma_{X(3872)}$ scan described in the text.
The region of the plot below the arrow contains 90\% of the total area
under the points. }
\label{fig:like}
\end{figure}

In the three-dimensional fit, there are correlations between the fitted width,
the numbers of signal  events ($n_{\rm sig}$) and  peaking background events ($n_{\rm peak}$). 
The other parameters have negligible correlations with the width.  We therefore 
performed a
series of fits to the data where we fixed $\Gamma_{X(3872)}$ at a sequence of values
ranging from 0.1 to 3.0~MeV.  In these fits all parameters other than $n_{\rm sig}$ and
$n_{\rm peak}$ were fixed at their best fit values; 
$n_{\rm sig}$ and $n_{\rm peak}$ were allowed to vary.   
Figure~\ref{fig:like} shows how the fit likelihood changes with $\Gamma_{X(3872)}$.
The arrow in the figure indicates the width  value, 
$\Gamma_{X(3872)} = 0.95$~MeV, below which 90\% of the integrated area under
the points is contained.   This value is below the experimental
resolution.  To check sensitivity to uncertainties in the
mass resolution width, we repeated the scan using
the value of the tail resolution width determined
from fitting the $\psip$ peak without any rescaling. 
This had negligible effect on the width of the likelihood.

In order to evaluate whether our measured limit is reasonable given the
size of our data sample,  
we derived width upper limits from similar analyses of 24 statistically independent,
170-event MC samples that were generated with $\Gamma_{X(3872)} = 0$. Of these,
twelve produced 90\% CL upper limits that are less than 1~MeV; five returned
a fit value at the lower limit imposed on the fit.  In a set of 24 MC samples
generated with $\Gamma_{X(3872)}=1$~MeV, none returned a width value at the 
lower limit of the fit and 17 produced 90\% CL {\it lower} limits that exclude zero.

The $\psip$ width has been precisely measured in $\ee$~\cite{BES2}
and $p\bar{p}$~\cite{E835}  threshold scans to be $0.304\pm 0.009$~MeV~\cite{PDG},
a value that is well below the resolution of our measurement.
We validated our experimental sensitivity to narrow natural widths
by refitting the $\psip$ data sample using resolution parameters fixed at
the values given in Table~\ref{tbl:psip} but 
with $\Gamma_{\psip}$ left as a free parameter.  The fit result
is $\Gamma_{\psip}=0.53 \pm 0.11$~MeV.  An examination of the fit
likelihood shows that it is well behaved and excludes a zero
width value by more than $4\sigma$.  The measured value
is $0.23 \pm 0.11$~MeV above the PDG's world-average value, which is
consistent with the bias value at $\Gamma\simeq 0.3$~MeV derived from the 
fitted curve in Fig.~\ref{fig:width_bias}, namely $0.25$~MeV.

As an upper limit on the natural width of the $X(3872)$, we inflate the
90\% CL value determined from the scan values shown in
Fig.~\ref{fig:like}  by $0.23$~MeV, the measured difference
between our measurement of $\Gamma_{\psip}$ and its world-average value,
to account for a possible measurement bias.  Since both the simulated
and observed biases are
positive and indicate that our measured limit is biased
high, this produces a conservative value for the upper limit.    
The result is
\begin{equation}
\Gamma_{X(3872)}< 1.2~{\rm MeV}~~~~~{\rm 90\% CL},
\end{equation}
which is more restrictive than the previous 90\% CL limit 
of 2.3~MeV~\cite{skchoi_x3872}.

\subsection{Product branching fractions}

We determine product branching fractions for 
$B^+\rt K^+ X$, $X\rt \pipi\jp$ and $B^0\rt K^0 X$,
$X\rt \pipi\jp$ via the relation
\begin{eqnarray}
{\mathcal B }(B\rt K X(3872))\times {\mathcal B }(X(3872)\rt\pipi\jp) &=&\nonumber
\\
\frac{ N^{\rm evts}_K} {N_{B\bar{B}}\epsilon_{K}  f_{K}
{\mathcal B}_{\jp\rt\ell\ell}},
\label{eq:branch_x3872}
\end{eqnarray}
\noindent
where the notation is the same as that used for Eq.~\ref{eq:branch_psip}.
The results are
\begin{eqnarray}
{\mathcal B}(B^+\rt K^+ X(3872))\times {\mathcal B}(X(3872)\rt\pipi\jp) & = &\nonumber
\\
(8.61\pm 0.82~{\rm (stat)} \pm 0.52~{\rm (syst)})\times 10^{-6},
\end{eqnarray}
and
\begin{eqnarray}
{\mathcal B}(B^0\rt K^0 X(3872))\times {\mathcal B}(X(3872)\rt\pipi\jp) & = & \nonumber
\\
(4.3\pm 1.2~{\rm (stat)} \pm 0.4~{\rm (syst)})\times 10^{-6},
\end{eqnarray}
where the systematic error includes uncertainties in the MC simulation
of the tracking, particle identification
for the leptons and charged kaon, $K_S$ reconstruction, uncertainties in the number of
$B\bar{B}$ meson pairs, choice of parameterization used in the three dimensional fit, 
MC statistics, decay model dependence and the error
on the world-average $\jp\rt\leplep$ branching fraction, all added in
quadrature.   The computations  
are summarized in Table~\ref{tbl:bf_syst}.  The ratio of the
$B^0$ and $B^+$ product branching fractions is
\begin{eqnarray}
R(X) = \frac {{\mathcal B}(B^0\rt K^0 X(3872))}
{{\mathcal B}(B^+\rt K^+ X(3872))} &= & \nonumber
\\
0.50\pm 0.14~{\rm (stat)} \pm 0.04~{\rm (syst)},
\end{eqnarray}
where the systematic error evaluation is summarized in Table~\ref{tbl:bf_syst}.
This value is above the range preferred by some molecular models for the $X(3872)$:
$0.06\le R(X) \le 0.29$~\cite{swanson_pr}.
The BaBar result for this ratio is $R(X) = 0.41 \pm 0.24 \pm 0.05$~\cite{babar_pipijpsi}.

\begin{table}[htb]
\begin{center}
\caption{\label{tbl:bf_syst}
Systematic errors on the product branching fraction measurement.}
\begin{tabular}{l|c|c|c}\hline
Source               &    $K^+X(3872)$    &   $K_S X(3872)$    &  $K_S/K^+$ Ratio      \\
                     &      (percent)     &    (percent)      &  (percent)            \\\hline\hline
$N_{B\bar{B}}$          &       1.4          &     1.4           &     -                 \\
Secondary BF        &       1.0          &     1.0           &     -                 \\
MC statistics        &       1.0          &     1.0           &    1.4                \\
MC model             &       2.1          &     2.1           &     -                 \\
Hadron ID            &       3.7          &     2.6           &    1.1                \\
Lepton ID            &       1.1          &     1.1           &     -                 \\
Tracking             &       1.8          &     1.4           &    0.4                \\
3-dim. fit model     &       3.0          &     5.0           &    6.0                \\
$K_S$ efficiency     &       -            &     4.5           &    4.5                \\ \hline
Quadrature sum       &       6.0          &     8.1           &    7.7                \\
\hline\hline
\end{tabular}
\end{center}
\end{table}

\section{Search for a charged partner of the $X(3872)$ in $B\rt K \pi^+\pi^0\jp$ decays}
We search for a charged partner of the $X(3872)$ decaying into
$\pi^+\pi^0\jp$ using the selection criteria described above for the $\pipi\jp$
analysis, with the exception that one of the charged pions is replaced by
a $\pi^0$.  For this we require two photons with $E_{\gamma}>35$~MeV
that reconstruct to a $\pi^0\rt\gamma\gamma$ with a mass-constrained fit $\chi^2\le 4.0$.
In the event of multiple $\gamma$ entries we choose the candidate with the best
$\chi^2$ from the $\pi^0$ mass-constrained fit; for multiple charged pions, 
we chose the candidate that produces the lowest value of $|\DE |$.  

\begin{figure}[htb]
\mbox{
  \includegraphics[height=0.25\textwidth,width=0.5\textwidth]{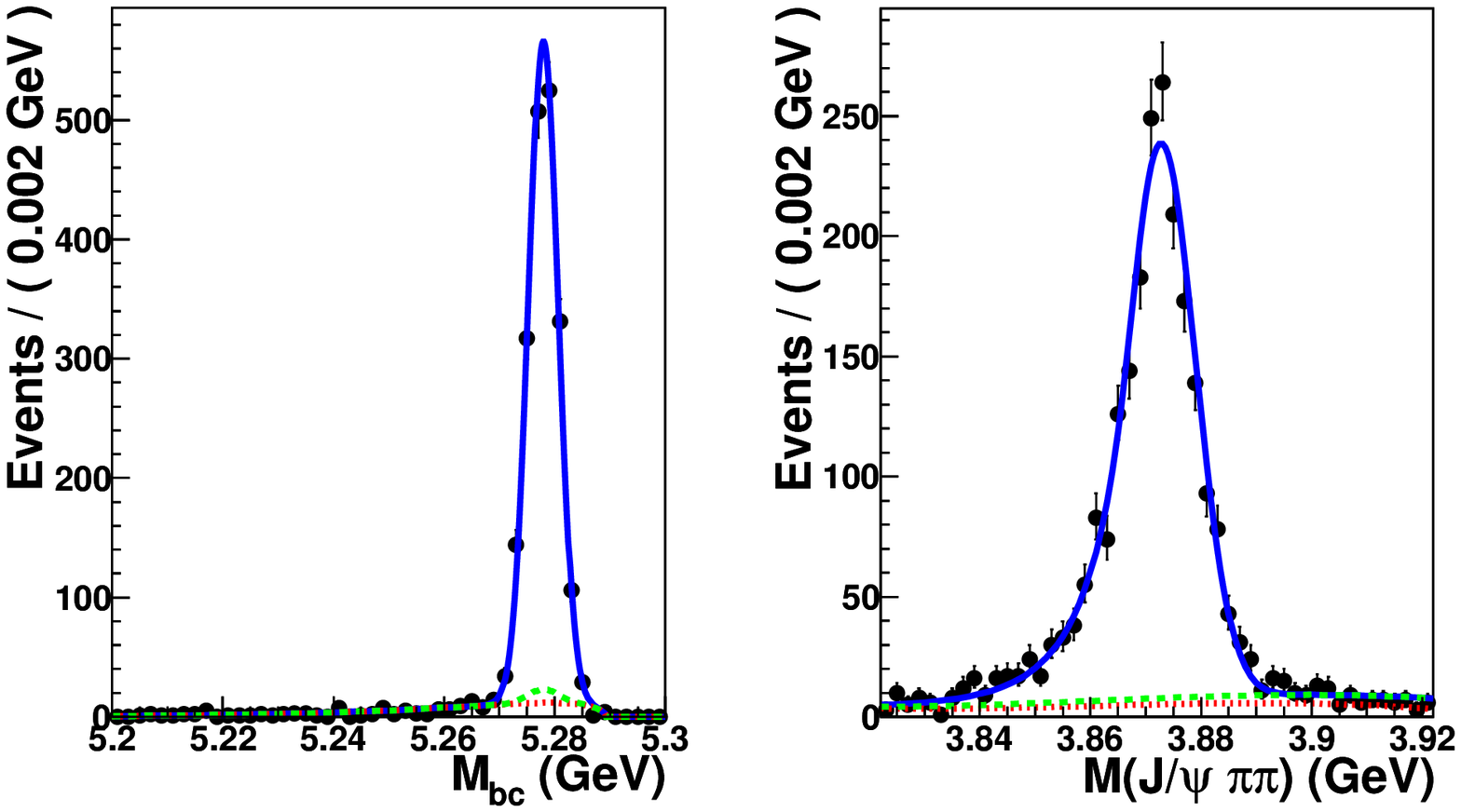}
}
\mbox{
  \includegraphics[height=0.25\textwidth,width=0.5\textwidth]{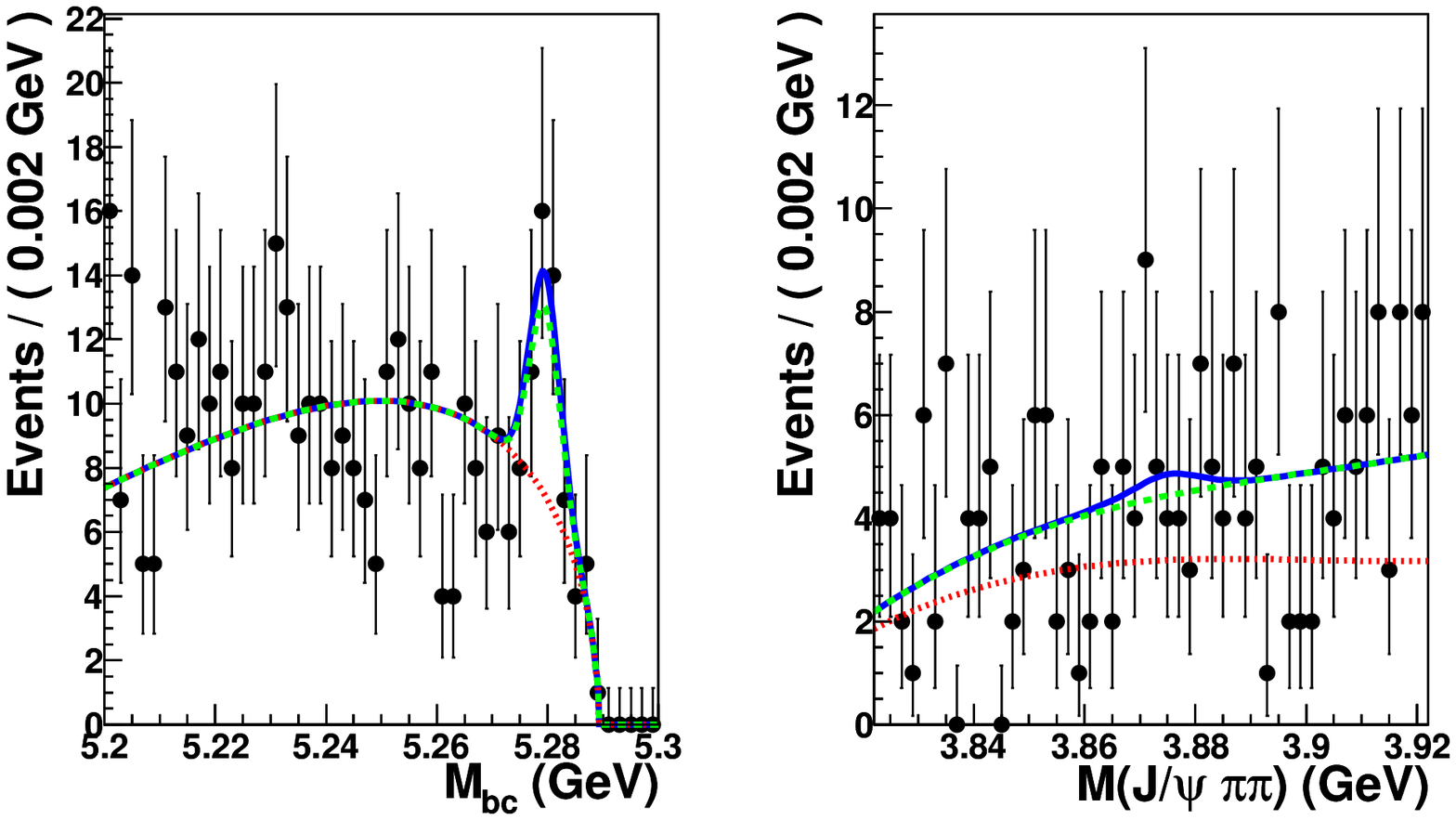}
}
\mbox{
  \includegraphics[height=0.25\textwidth,width=0.5\textwidth]{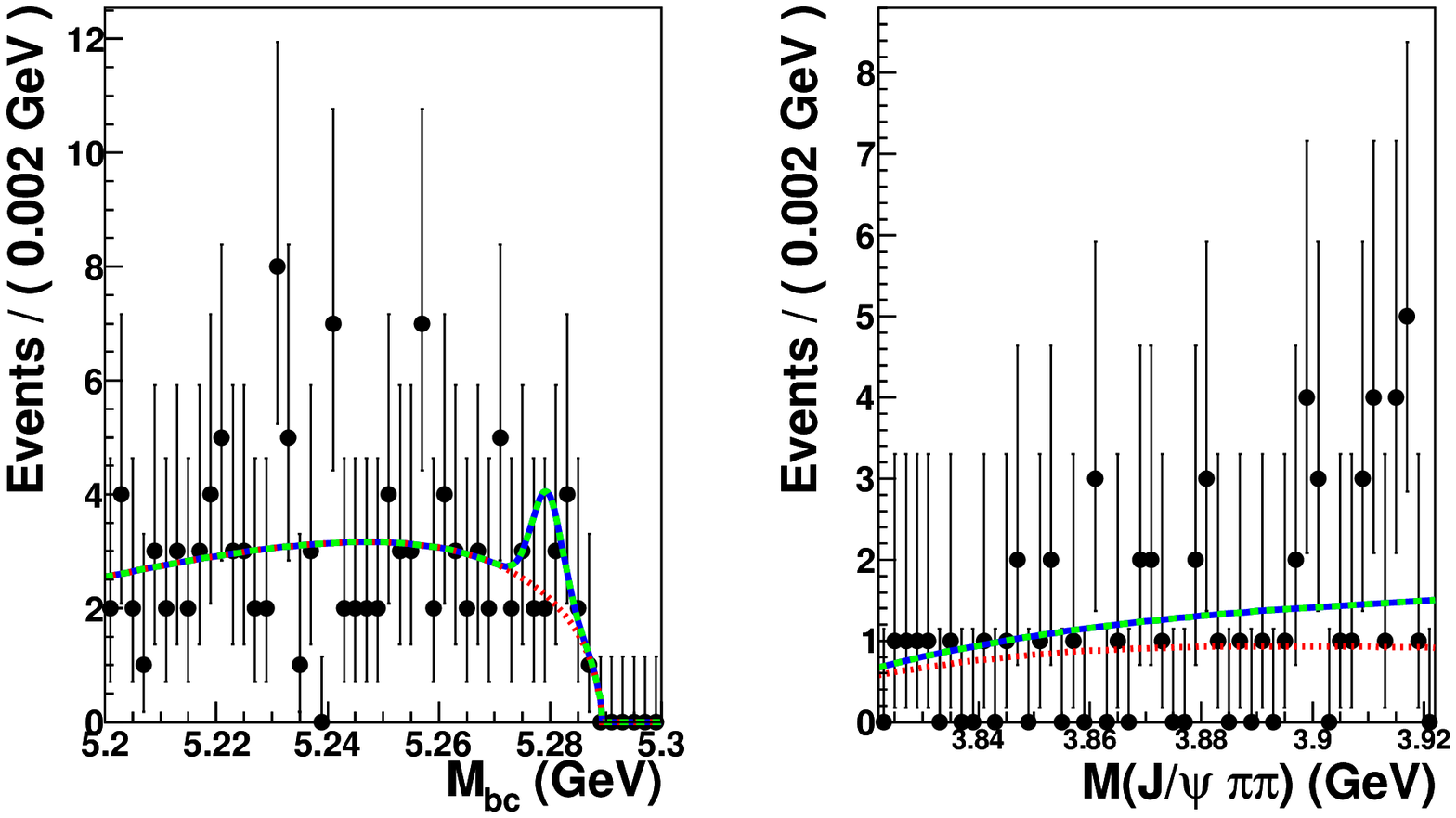}
}
\caption{ 
The $\Mbc$ (left), $M(\pi^+\pi^0\jp)$ (right) distributions for  
$B\rt K X^+(3872)$, $X^+\rt\rho^+\jp$ MC events (top) and 
$\bar{B}^0 \rt K^- \pi^+\pi^0\jp$ (middle) and $B^+ \rt K^0 \pi^+\pi^0\jp$ (bottom)
event candidates in the data, within the signal 
region of the other quantity.   The curves show
the results of the fits described in the text.
}
\label{fig:mpipi0jpsi_kx3872_3box}
\end{figure}

We perform an unbinned two-dimensional ($\Mbc$ {\it vs.} $M(\pi^+\pi^0\jp)$) maximum likelihood
fit to the selected event samples using Gaussian and ARGUS function PDFs for the $\Mbc$
signal and background, and a Crystal Ball function~\cite{Xtal_ball} and third-order polynomial for the
$M(\pi^+\pi^0\jp)$ signal and background, respectively.  For the peaking background
we use the $\Mbc$ signal PDF and a linear background shape for the  $M(\pi^+\pi^0\jp)$ PDF.
The Crystal Ball function parameters are fixed at values returned from fits to 
samples of Monte Carlo simulated $B\rt K X^+$, $X^+\rt \rho^+\jp $ events with 
$m_{X^+} =3871.7$~MeV and $\Gamma_{X^+} = 0$.  
The results of the fit to the
simulated $\bar{B}^0\rt K^-X^+$ sample are 
shown in the top panels of Fig.~\ref{fig:mpipi0jpsi_kx3872_3box}. 

For the data, we do a series of fits with the $X^+$ mass restricted to
overlapping 10~MeV mass windows covering the range 3850~MeV to 3890~MeV.
For the $K^-X^+$ channel the largest signal yield is $4.2\pm 7.8$ events 
at a mass of $3873\pm 6$~MeV.   The 90\% CL upper limit, corresponding to
the signal yield below which 90\% of the area of the likelihood function is contained,  
is 17.3 events.  For the $K^0X^+$ channel, all mass intervals
have a zero signal yield and the 90\% upper limit derived from the likelihood 
function for a peak mass fixed at $3873$~MeV is 5.4 events.
$\Mbc$ and $M(\pi^+\pi^0\jp )$ plots for the fit to the $K^-X^+$ sample with the
highest event yield are shown in the middle panel of
Fig.~\ref{fig:mpipi0jpsi_kx3872_3box}.  The bottom panels of 
Fig.~\ref{fig:mpipi0jpsi_kx3872_3box}
show the results of the fit to the $K^0X^+$ sample with peak mass fixed at 3873~MeV.

We determine 90\% CL product branching fraction upper limits using the relation
 \begin{equation}
{\mathcal B }(B\rt K X^+)\times {\mathcal B }(X^+\rt\rho^+\jp) <
\frac{ N^{\rm evts}_{\rm 90\% UL}} {N_{B\bar{B}}{\mathcal B}_{\jp\rt\ell\ell}
\epsilon_{K}  f_{K}} ,
\end{equation}
where $ N^{\rm evts}_{\rm 90\% UL} $ is the upper limit on
the event yield for each channel, 
$f_{K^+}=1.0$ and $f_{K^0}=0.346$ (as in Eq.~\ref{eq:branch_psip}),
and $\epsilon_{K}$ are the MC acceptances 
reduced by the systematic error: $\epsilon_{K+} = 4.5\%$ and 
$\epsilon_{K^0}= 2.8 \%$.  
The systematic errors are the
same as those listed in Table~\ref{tbl:bf_syst} above, with the additional
inclusion of a 3\% systematic error associated
with data-MC differences in $\pi^0$ detection and
2.5\% for the increase in the upper bounds when the resolution parameters
of the $M(\pi^+\pi^0\jp)$ signal PDF are varied by $\pm 10\%$.  
The systematic errors are 6\% for the
$B^0\rt K^-X^+$ and 8\% for the $B^+\rt K^0X^+$ channels.
The resulting limits are
\begin{equation}
{\mathcal B}(\bar{B}^0\rt K^- X^+)\times {\mathcal B}(X^+\rt\rho^+\jp) < 4.2 \times 10^{-6}
\end{equation}
and
\begin{equation}
{\mathcal B}(B^+\rt K^0 X^+)\times {\mathcal B}(X^+\rt\rho^+\jp) < 6.1 \times 10^{-6}.
\end{equation}
The BaBar limits for the same quantities are
${\mathcal B}(\bar{B}^0\rt K^- X^+)\times {\mathcal B}(X^+\rt\rho^+\jp) < 5.4 \times 10^{-6}$
and
${\mathcal B}(B^+\rt K^0 X^+)\times {\mathcal B}(X^+\rt\rho^+\jp) < 22 \times 10^{-6}$~\cite{babar_pipi0jpsi}.

\section{Angular correlation studies}

For subsequent analysis,
we define a tighter $X(3872)$ signal region that extends
$\pm 6$~MeV around the $M(\pipi\jp)$ signal peak. For
background estimates we use $\pm 12$~MeV sidebands 
above and below the signal peak
centered at 3852~MeV and 3892~MeV. 
There are in total 165 events 
in the signal region;  the background content, 
determined from the scaled sidebands, is $34 \pm 3$ events.

Angular distributions for the sequential
decays $B\rt K X(3872)$, $X(3872)\rt\rho\jp$, $\rho\rt\pipi$
and $\jp\rt\leplep$ for the $1^{++}$ and $2^{-+}$
cases are given by the LHCb group in Ref.~\cite{LHCb-PUB-2010-003}.  
Since both the $B$ and $K$ mesons are scalar particles,
an $X(3872)$ meson produced via exclusive $B\rt KX$ decays
must have a zero component of angular momentum along 
its  momentum direction in the $B$ rest frame and, thus, its
polarization vector, $\vec{\epsilon}_X$, must be along this
boost direction.  This limits the number of
independent partial-wave amplitudes needed to describe the 
decay.   Moreover, angular momentum and parity conservation
in $X(3872)\rt\rho\jp$ decay implies that for $1^{++}$ the $\rho$
and $\jp$ are in an $S$- and/or $D$-wave, while for $2^{-+}$
they are in a $P$- and/or $F$-wave.  Since the 
$X(3872)\rt \rho\jp$ decay occurs at threshold, only
the lower partial wave in each case is considered.
With this constraint, the $1^{++}$ has only one
decay amplitude: $L=0$ and $S=1$, where $L$ the $\rho$-$\jp$ orbital angular
momentum and $S$ their spin state.
The $2^{-+}$ hypothesis has two independent amplitudes: $L=1$ with
$S=1$ or $S=2$, which we denote by $B_{11}$ and $B_{12}$, respectively.

\begin{figure}[htb]
\mbox{
  \includegraphics[height=0.25\textwidth,width=0.4\textwidth]{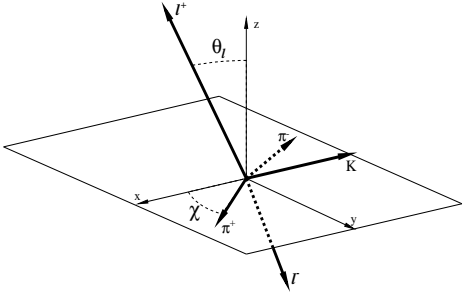}
}
\caption{ 
Definitions of the angles $\chi$ and $\theta_{\ell}$ 
as described in the text.
}
\label{fig:helicity}
\end{figure}

We denote by $\theta_X$ the angle between the $\jp$ and the direction
opposite to the kaon in the $X(3872)$ restframe.  
In the case of $J^{PC}=1^{++}$, the $X(3872)\rt\rho\jp$ decay produces
a $\rho$ and $\jp$  in an $S$-wave and, thus, the distribution in $\cos\theta_X$ is
expected to be flat.
For $2^{-+}$, the final state is $P$-wave and the $\cos\theta_X$ distribution
is $\propto (1+3\cos^2\theta_X)$ for $B_{12}=0$,  approximately flat for 
$|B_{11}| \simeq |B_{12}|$,  and $\propto \sin^2\theta_X$ for $B_{11}=0$.  
For $1^{++}$ decays to an $S$-wave at threshold, the interaction 
Lagrangian is $
{\mathcal L}_{\rm int} \propto\vec{\epsilon}_X\cdot(\vec{\epsilon}_{\jp}\times\vec{\epsilon}_{\rho})$,
where $\vec{\epsilon}_{\jp}$ and $\vec{\epsilon}_{\rho}$ polarization vectors.
Thus, the three polarization vectors tend to be mutually perpendicular.
In polarized $\rho\rt\pipi$ decays,  the pions have a $\cos^2\theta$ distribution relative to
the $\vec{\epsilon}_{\rho}$ direction, while in  polarized $\jp\rt\leplep$, the decay leptons have a 
$\sin^2\theta$ distribution relative to the $\vec{\epsilon}_{\jp}$ direction.  To exploit this,
we use a coordinate system suggested by Rosner~\cite{rosner} where the $x$-axis is the direction
opposite to the kaon ({\it i.e.}, the $\vec{\epsilon}_X$ direction), 
the $x-y$ plane is defined by the kaon and $\pi^+$ and the $z$ axis completes
a right-handed coordinate system.  The angle between the $\pi^+$ direction
and the $x$-axis is designated as $\chi$ and the angle  between the $\ell^+$ 
direction and the $z$-axis as $\theta_{\ell}$,
as shown in Fig.~\ref{fig:helicity}.  In the 
limit where the $\jp$ and $\rho$ are at rest in the $X$ rest frame,
the expectation for $1^{++}$ has the distinctive pattern
\begin{equation}
\frac{d^2N}{d\cos\theta_{\ell}d\cos\chi}
\propto \sin^2\theta_{\ell} \sin^2\chi .
\end{equation}
The changes in the values of $\cos\chi$ and $\cos\theta_{\ell}$ that occur when 
$\chi$ and $\theta_{\ell}$ are determined in either the $\jp$ or $\rho$ restframes 
(instead of the $X(3872)$ frame) are much smaller than the bin
sizes used in this analysis.

The CDF results on angular correlations used a three-dimensional fit to
data divided into twelve bins~\cite{CDF_angles}.  The limited statistics of our sample
preclude dividing the data into enough bins to make a three-dimensional
fit feasible.  Instead we compare one-dimensional histograms of data and
MC for different hypotheses.

\begin{figure}[htbp]
\mbox{
  \includegraphics[height=0.24\textwidth,width=0.48\textwidth]{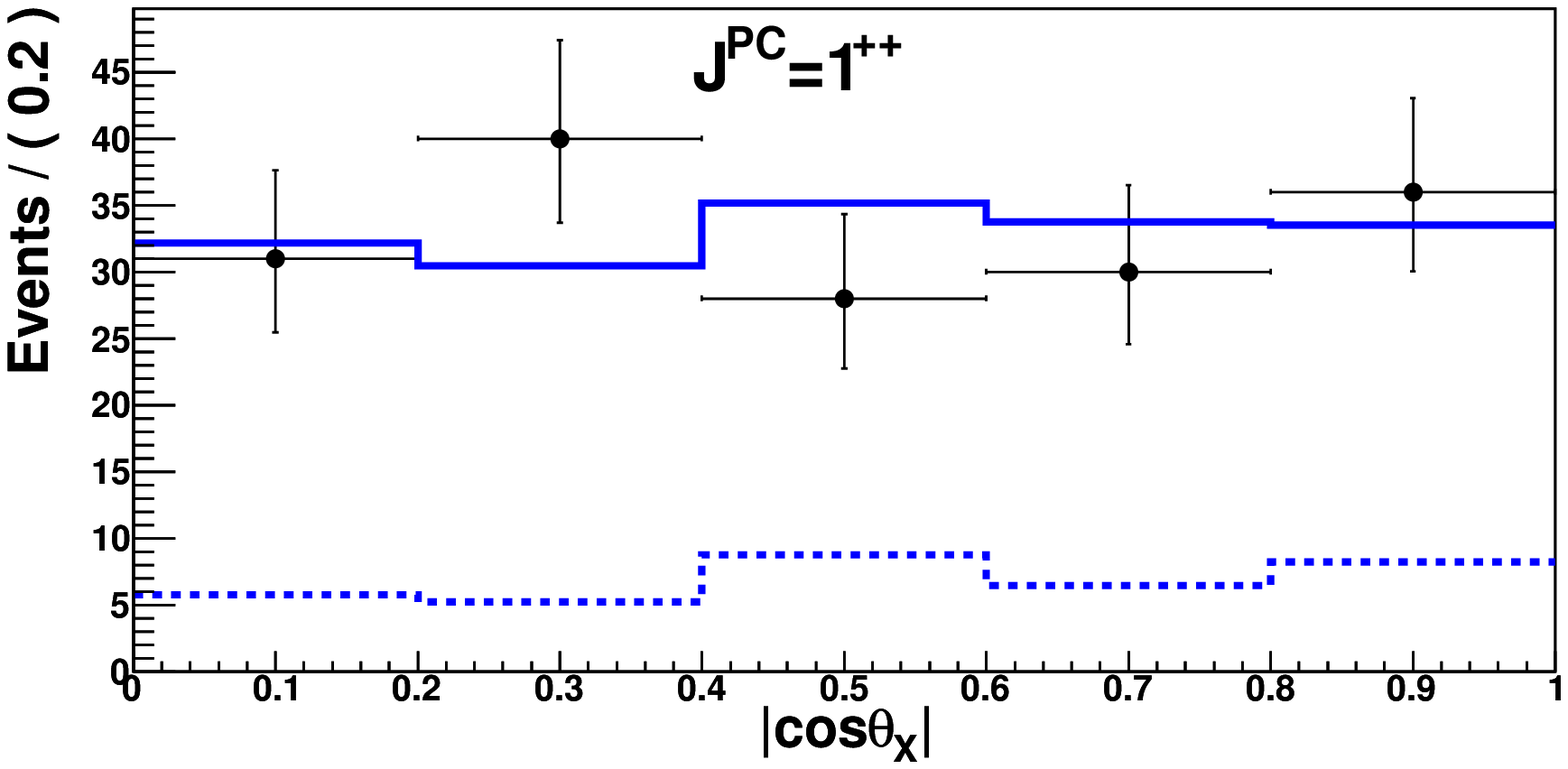}
}
\mbox{
  \includegraphics[height=0.24\textwidth,width=0.48\textwidth]{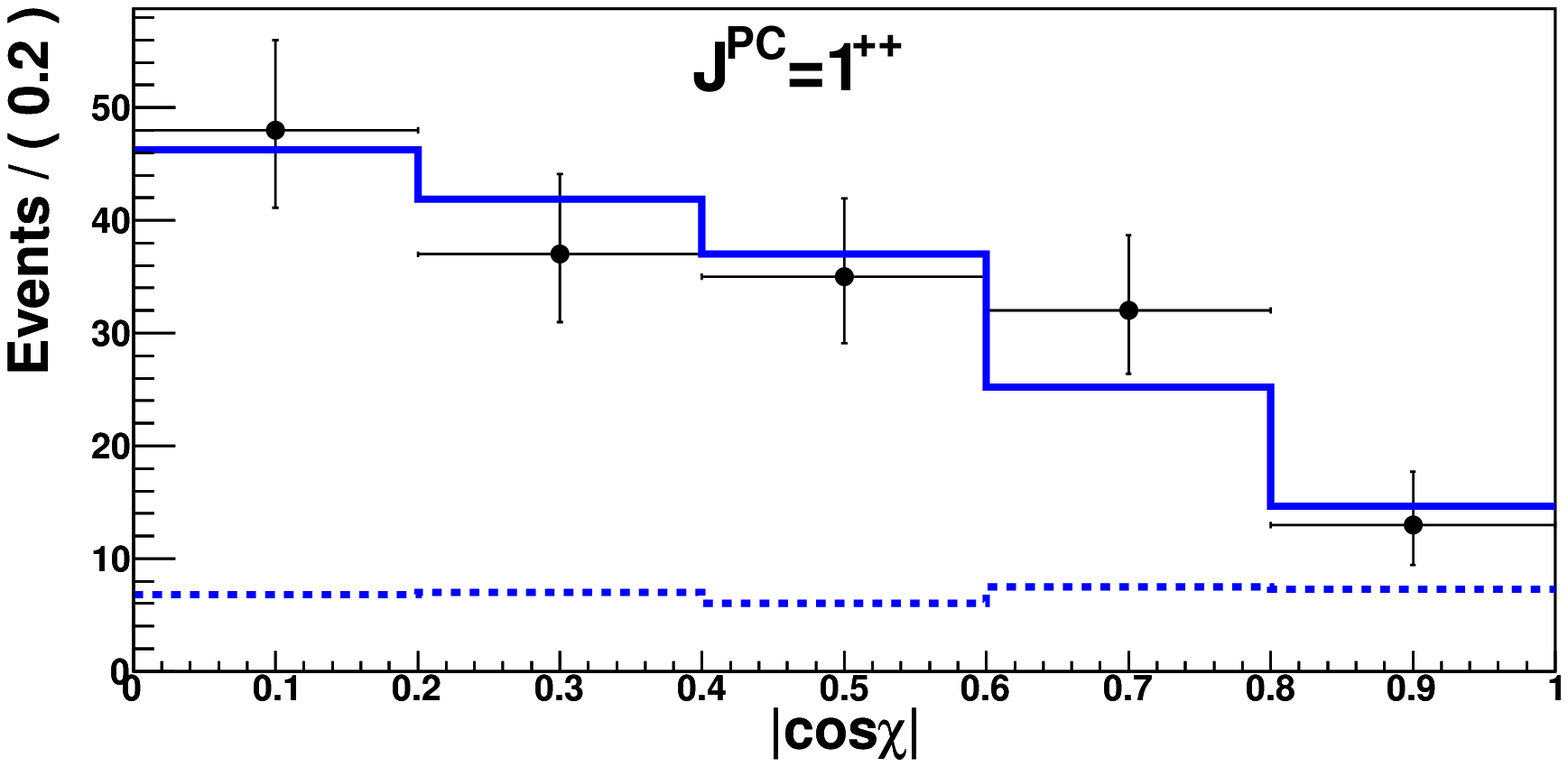}
}
\mbox{
  \includegraphics[height=0.24\textwidth,width=0.48\textwidth]{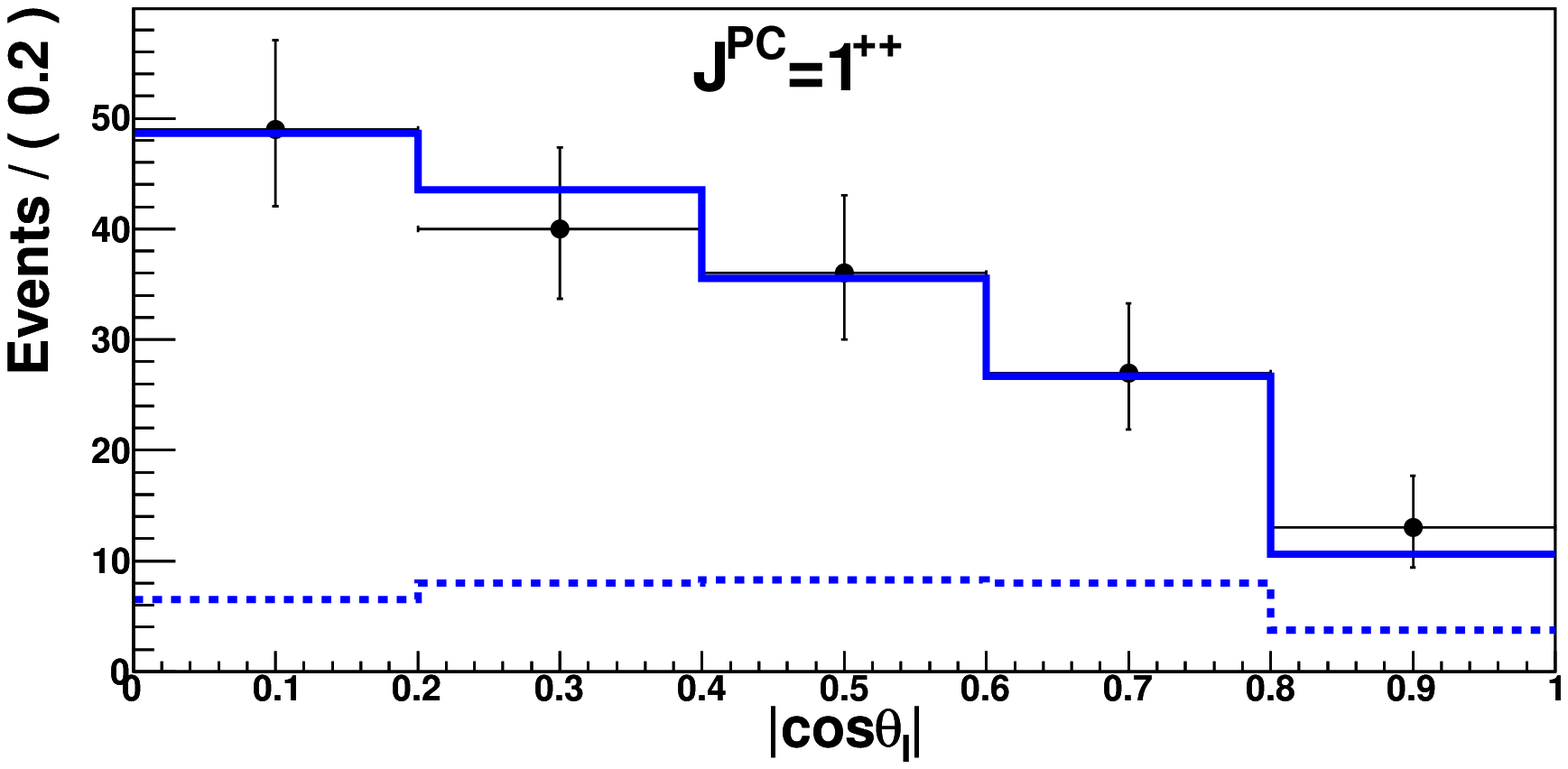}
}
\caption{The comparisons described in the text for the $J^{PC}=1^{++}$ hypothesis
applied to $|\cos\theta_X|$ (top), $|\cos\chi|$ (middle) and $|\cos\theta_{\ell}|$
(bottom).  The dashed histograms indicate the sideband-determined background levels.
}

\label{fig:angles_1++_fit}

\end{figure}

The data points in Fig.~\ref{fig:angles_1++_fit}
show the 
$|\cos\chi|$, $|\cos\theta_{\ell}|$ and
$|\cos\theta_X|$ distribution for
$X(3872)$ signal region events. The dotted
histograms indicate the background determined  
from the events in the scaled $M(\pipi\jp)$
sidebands.  The solid histogram is the sum of
the background (dotted histogram) and simulated 
MC $X(3872)\rt\rho\jp$ events generated with a 
$1^{++}$ ($S$-wave only) hypothesis and
normalized to the observed signal.
(The MC samples described in this section were generated 
using the partial wave option of EvtGen~\cite{evtgen}.)  
With no other free parameters, we find 
good matches between $1^{++}$ expectations and 
the data for all three distributions:
the $\chi^2$ values (confidence levels) are 
$3.82$ (0.43), $1.76$ (0.78) and $0.56$ (0.97)
for $|\cos\theta_X|$, $|\cos\chi|$ and 
$|\cos\theta_{\ell}|$, respectively.

For $J^{PC}=2^{-+}$, in addition to the normalization, there are two more free parameters
that we take to be the ratio $|B_{11}|/|B_{12}|$ and the relative phase
between $B_{11}$ and $B_{12}$. 
A comparison of the measured distributions with those for a MC simulated $2^{-+}$
state with  $B_{11} = 0$ finds poor matches for all three
angular distributions: the $\chi^2$ values (confidence levels) are 
$14.9$ (0.005), $48.8$ ($<10^{-7}$) and $16.5$ (0.002)
for $|\cos\theta_X|$, $|\cos\chi|$ and $|\cos\theta_{\ell}|$, respectively.
For $B_{12}=0$, 
there are reasonable matches between data and MC
for the $|\cos\chi|$ ($\chi^2 = 6.04$, CL=0.20) and $|\cos\theta_{\ell}|$
($\chi^2=1.92$, CL=0.75) distribution,
but poor agreement in the case of the $|\cos\theta_X|$
comparison ($\chi^2=16.2$, CL=0.003).  

\begin{figure}[htbp]
\mbox{
  \includegraphics[height=0.24\textwidth,width=0.48\textwidth]{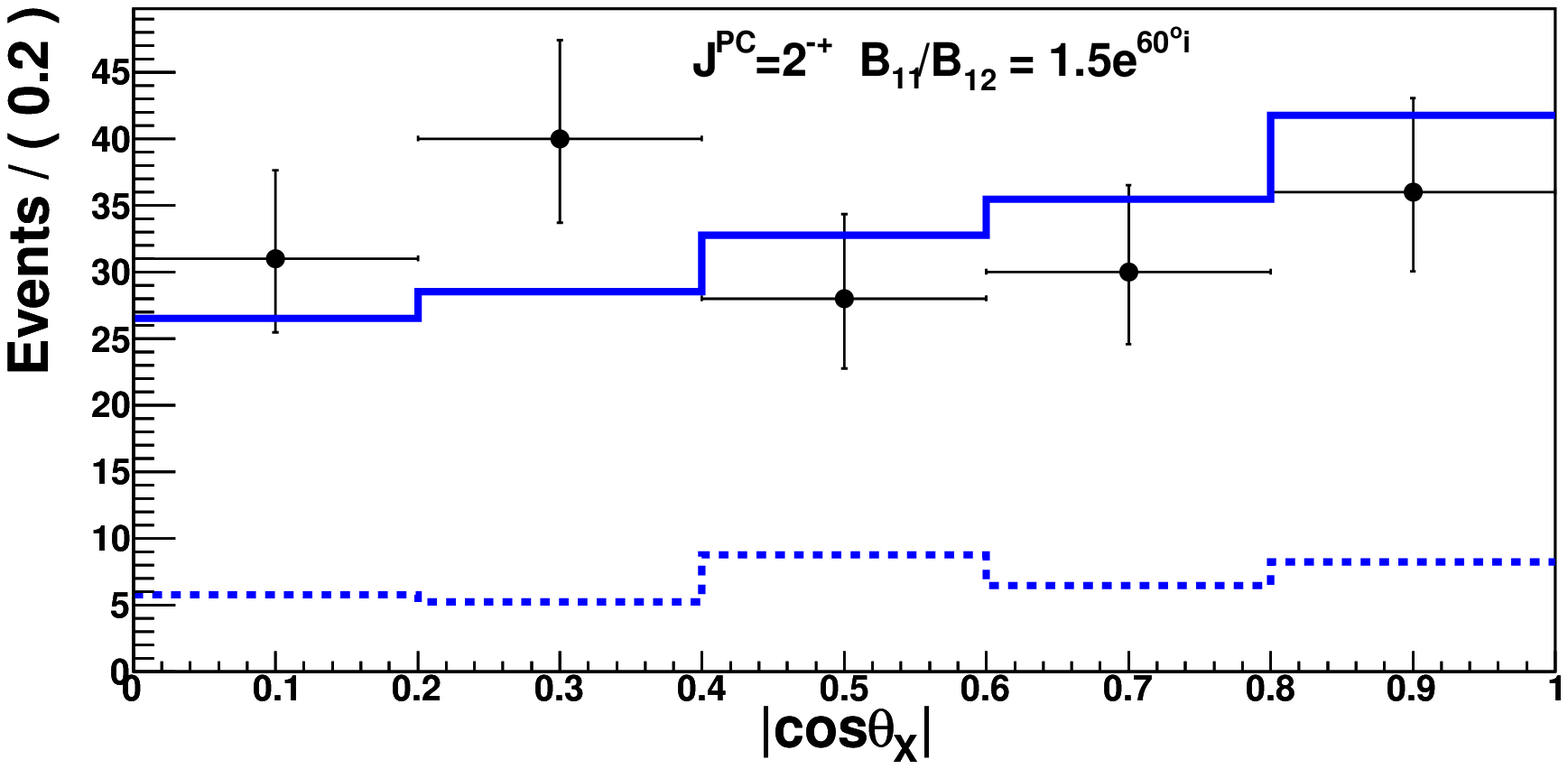}
}
\mbox{
  \includegraphics[height=0.24\textwidth,width=0.48\textwidth]{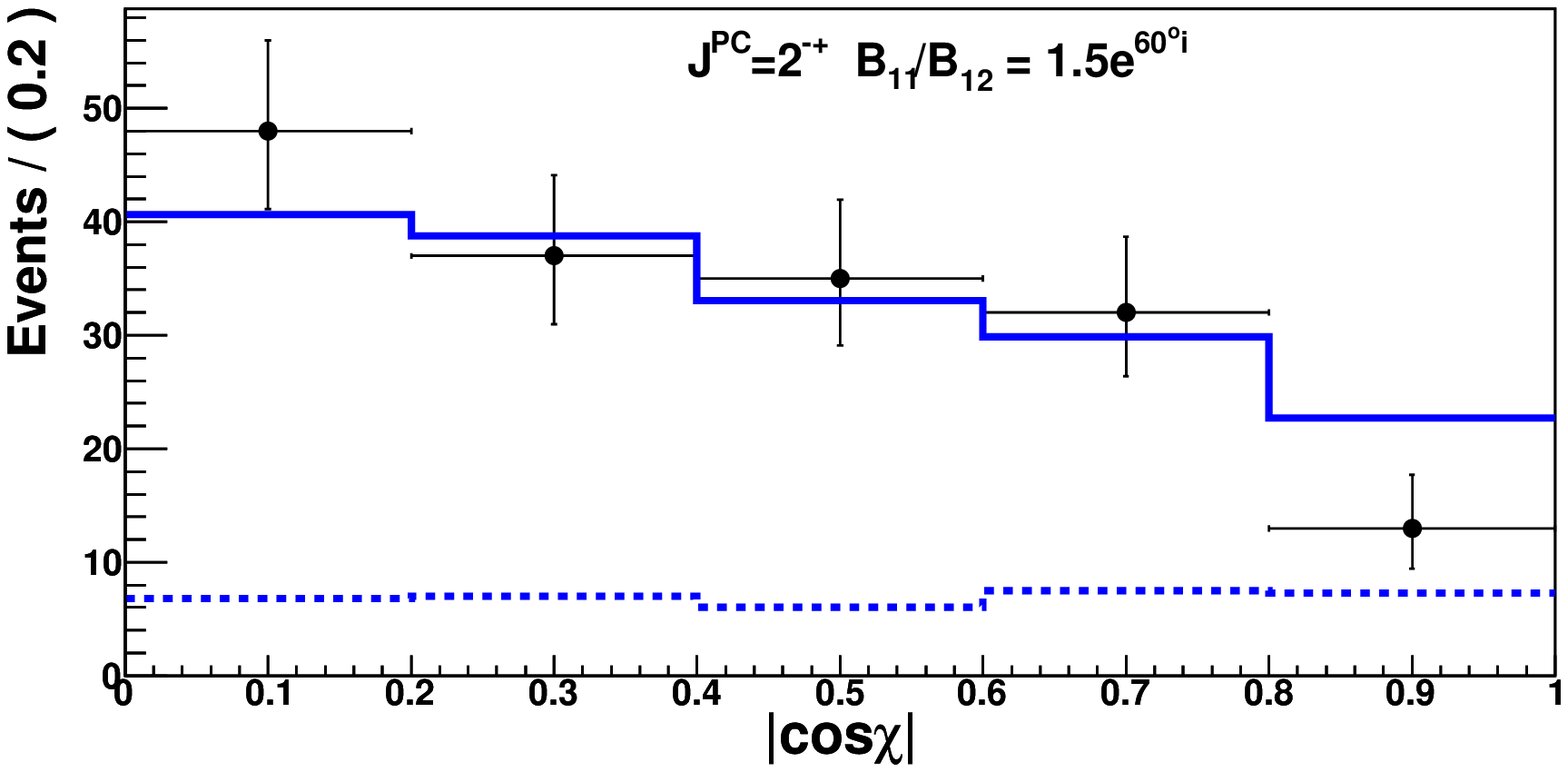}
}
\mbox{
  \includegraphics[height=0.24\textwidth,width=0.48\textwidth]{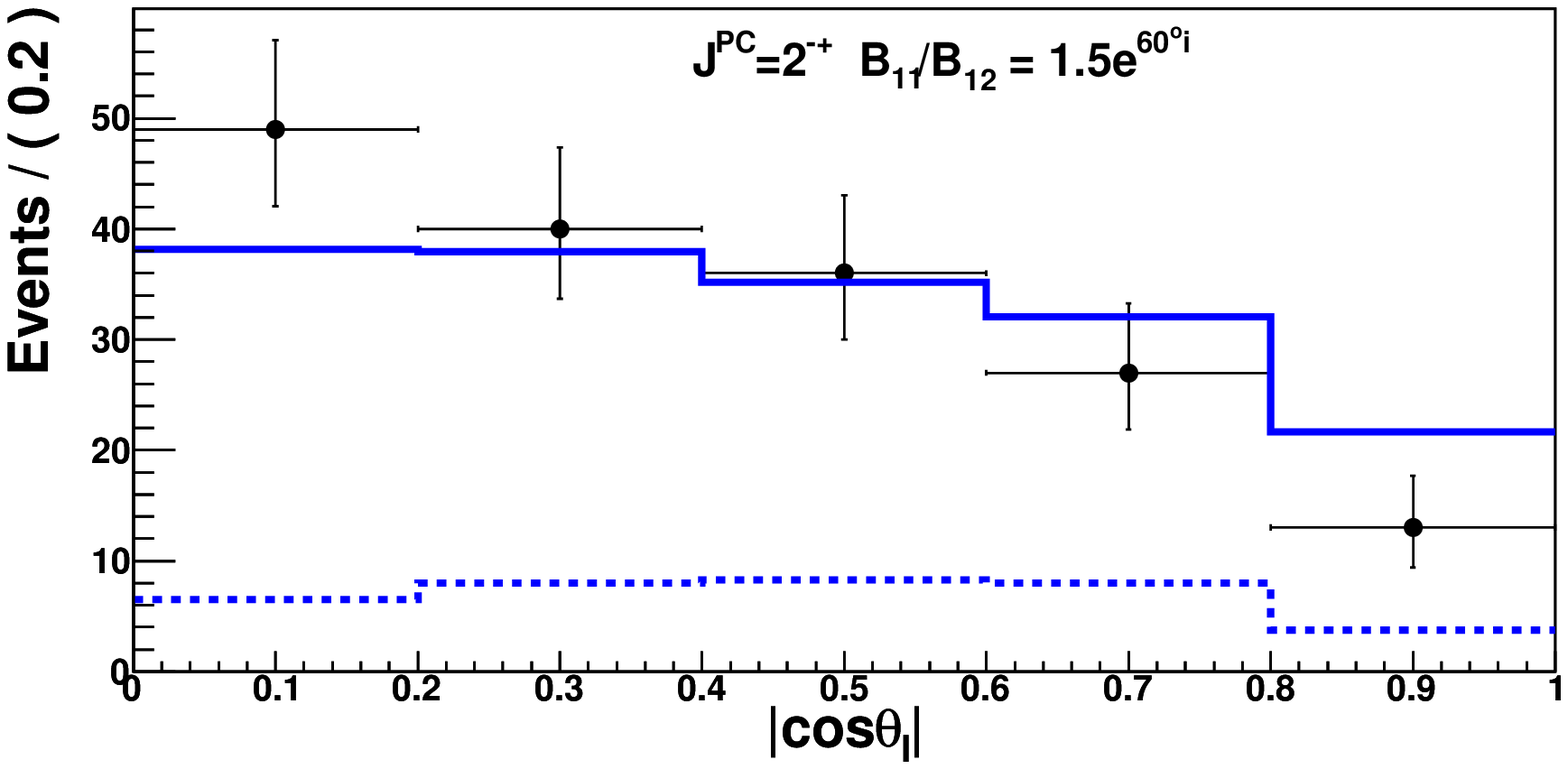}
}
\caption{The comparisons described in the text for the $J^{PC}=2^{-+}$ hypothesis
applied to $|\cos\theta_X|$ (top), $|\cos\chi|$ (middle) and $|\cos\theta_{\ell}|$
(bottom) for $B_{11}/B_{12}=1.5e^{60^{\circ}i}$. 
}

\label{fig:angles_2-+_b_fit}

\end{figure}

We made similar comparisons with simulated event samples
for a grid of values for 
$|B_{11}|/|B_{12}|$ and its relative phase.  
Figure~\ref{fig:angles_2-+_b_fit} shows the data~-~MC 
comparison for the case where $(B_{11}/B_{12}) = 1.5e^{60^{\circ}i}$,
the value for which we found the best match.  
In this case all three MC distributions have acceptable 
$\chi^2$ values (confidence levels): 
$4.72$ (0.32) for $\cos\theta_X$, $4.60$ (0.33) for $\cos\chi$,
and $5.24$ (0.26) for $\cos\theta_{\ell}$.  
The LHCb analysis uses the parameter
$\alpha=\frac{B_{11}}{B_{11}+B_{12}}$~\cite{LHCb-PUB-2010-003};
the values of $|B_{11}|/|B_{12}|$ and the relative phase that are
listed above translate into $\alpha = 0.69 e^{23^{\circ}i}$.

We conclude that with the
current level of statistical precision we cannot
distinguish definitively between the $1^{++}$
and $2^{-+}$ assignments.  However, while the $2^{-+}$
MC distributions for all three angles are
similar to  those for $1^{++}$, they
differ in detail, suggesting that in future experiments with larger data
samples, such as LHCb~\cite{LHCb-detector}, Belle II~\cite{belle2_exp}
and SuperB~\cite{superb},
three-dimensional fits based on the angles discussed here
will be able to distinguish between the two $J^{PC}$ hypotheses.

\section{Fits to the $M(\pipi)$ distribution}

For even-parity $C=+1$ states 
the $\pipi\jp$ final state would be a  $\rho$ and $\jp$ primarily 
in a relative $S$-wave, while for $2^{-+}$,  
the $\rho$ and $\jp$ would be in a relative $P$-wave.  
For the $S$-wave case, the 
$M(\pipi)$ mass distribution near the upper kinematic limit 
is modulated by the available phase space, which is proportional
to $k^*$, the $\jp$ momentum in the $X(3872)$ restframe. 
For a $\jp$ and $\rho$ in a $P$-wave, the upper boundary is 
suppressed by an additional $(k^{*})^{2}$ centrifugal barrier.
Thus, the high-mass 
part of the $\pipi$ invariant mass distribution
provides some $J^{P}$ information.

\begin{figure}[htbp]
\mbox{
  \includegraphics[height=0.35\textwidth,width=0.5\textwidth]{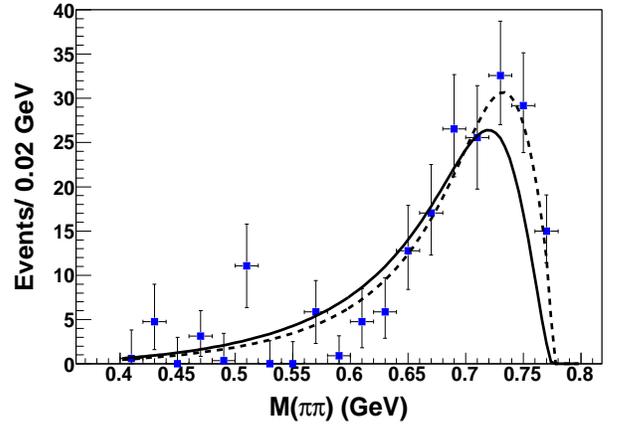}
}
\caption{The data points show the background-subtracted,
relative-efficiency-corrected $M(\pipi)$  distribution 
for $X(3872)\rt\pipi\jp$ events. The curves
show the results of fits using an $S$-Wave (dashed) and a $P$-Wave (solid)
BW function as described in the text.}

\label{fig:mpipi_swave_pwave_2box}

\end{figure}

We extract a background-subtracted $M(\pipi)$ spectrum from 
a series of two-dimensional ($\Mbc$ {\it vs.} $\DE$) likelihood
fits to data in 20~MeV-wide $M(\pipi)$ bins covering the range 
$0.4~{\rm GeV}\le M(\pipi)\le 0.78$~GeV.
The extracted yields are corrected for the $M(\pipi)$-dependence
of the experimental acceptance using results from simulated data
samples of $B\rt K X$, $X\rt \rho\jp$ events where the $\rho$ mass is 
artificially set at various masses with a narrow width. 
The peaking background remaining in the data is estimated
from the $M(\pipi\jp)$ sidebands to be $12\pm 5$ events with an 
$M(\pipi)$ distribution that is similar
to that of the $X(3872)$ signal.    The 
resulting distribution is shown as data points with error bars in
Fig.~\ref{fig:mpipi_swave_pwave_2box}

We fit the $M(\pipi)$ distribution for events in the 
$X(3872)$ signal region using the parameterization of Ref.~\cite{cdf_pipi}
\begin{equation}
dN/dm_{\pi\pi} \propto (k^*)^{2\ell+1}f^2_{\ell X}(k^*)|BW_{\rho}(m_{\pi\pi})|^2,
\label{eq:mpipi_form}
\end{equation}
where $k^*$ is defined above, $\ell$ is the orbital angular momentum value,
$f_{0X}=1.0$ and $f_{1 X}(k^*)=(1+R_X^2k^{*2})^{-1/2}$ are Blatt-Weisskopf 
``barrier factors''~\cite{bw} and $BW_{\rho}$ is the relativistic
BW expression
\begin{equation}
BW_{\rho}(m_{\pi\pi}) \propto \frac{\sqrt{m_{\pi\pi}\Gamma_{\rho}}}
{m^2_{\rho} - m^2_{\pi\pi} - im_{\rho}\Gamma_{\rho}}.
\label{eq:relBW}
\end{equation}
Here $\Gamma_{\rho}=\Gamma_{0}[q^*/q_{0}]^{3}[m_{\rho}/m_{\pi\pi}][f_{1\rho}(q^*)/f_{1\rho}(q_0)]^2$,
where $q^*(m_{\pi\pi})$ is the pion momentum in the $\rho$ rest frame,
$q_0=q^*(m_{\rho})$, $f_{1\rho}(q))=(1+R^2_{\rho}q^2)^{-1/2}$,
$\Gamma_{0}=146.2$~MeV and $m_{\rho}=775.5$~MeV~\cite{PDG}.
The ``radii'' $R_X$ and $R_{\rho}$ are poorly known.
Generally $R_{\rho}=1.5$~GeV$^{-1}$ is used and CDF uses values for $R_X$
that are as large as $R_X=5.0$~GeV$^{-1}$.  (Higher values of $R_X$ reduce
the effects of the $k^{*(2\ell + 1)}$ factor and, therefore, make the $S$- and $P$-wave
differences smaller.)   We take these values as our default settings.   

The smooth curves in Fig.~\ref{fig:mpipi_swave_pwave_2box} show the results of the
$S$-wave (dashed line)  and $P$-wave (solid line) fits.  
The $S$-wave ($\ell=0$) case fits the data well: $\chi^2/d.o.f. = 17.5/18$ (CL=49\%).
The $P$-wave ($\ell=1$) fit is poorer, $\chi^2/d.o.f. = 32.1/18$ (CL=2\%). 
Reducing the Blatt-Weisskopf radius for the $X(3872)$ makes the $P$-wave fit worse,
increasing  $R_X$ to $7.0$~GeV$^{-1}$ improves
the $P$-wave fit $\chi^2/d.o.f.$ to $26.5/18$, which
corresponds to a 9.0\% CL.  Large changes in $R_{\rho}$ are found to 
have little effect on the fit quality for either case.

However, both Belle~\cite{belle_3pijpsi} and BaBar~\cite{babar_3pijpsi} have
reported evidence for the sub-threshold decay process $X(3872)\rt\omega\jp$.
The CDF group pointed out that interference between the $\rho\jpsi$ and
$\omega\jpsi$ final states, where $\omega\rt\pipi$, can have important 
effect on the $M(\pipi)$ lineshape near the upper kinematic limit~\cite{cdf_pipi}.
We therefore repeated the above-described fits with the inclusion of possible
effects from $\rho$-$\omega$ interference.

For these fits we use the form given in Eq.~\ref{eq:mpipi_form} with $BW_{\rho}(m_{\pi\pi})$
replaced by
\begin{equation}
BW_{\rho -\omega} \propto BW_{\rho} + r_{\omega}e^{i\phi_{\omega}} BW_{\omega},
\label{eq:rwint}
\end{equation}
where $BW_{\omega}$ is the same form as $BW_{\rho}$ with $\omega$ meson mass and width
values substituted for those of the $\rho$, $r_{\omega}$ is the strength of
the $\omega$ amplitude relative to that of the $\rho$, and $\phi_{\omega}$ is their relative
phase, which is expected to be $95^{\circ}$~\cite{rw_phase}.  

\begin{figure}[htbp]
\mbox{
  \includegraphics[height=0.35\textwidth,width=0.5\textwidth]{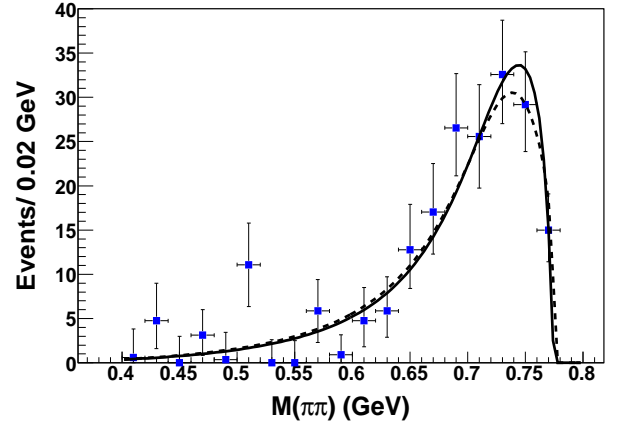}
}
\caption{The background-subtracted, relative-efficiency-corrected $M(\pipi)$  distribution 
for $X(3872)\rt\pipi\jp$ events. The curves
show the results of fits using an $S$-Wave (dashed line) and a $P$-Wave (solid line)
BW function with effects of $\rho$-$\omega$ interference included.  
}

\label{fig:mpipi_swave_pwave_rwint_2box}

\end{figure}

We performed fits to the 
$M(\pipi)$ distribution using this form weighted by the acceptance
with $\phi_{\omega}$ fixed at $95^{\circ}$ and $r_{\omega}$
left as a free parameter.   Figure~\ref{fig:mpipi_swave_pwave_rwint_2box} shows the results of
the $S$-wave (dashed line) and $P$-wave (solid line) fits.  
The inclusion of a small $\omega$ amplitude ($r_{\omega}=0.07 \pm 0.05$) improves the
$S$-wave fit to $\chi^2/d.o.f.=15.8/17$ (54\% CL).  The $P$-wave fit returns a 
larger $\omega$ contribution, $r_{\omega}=0.48^{+0.20}_{-0.14}$, and a good fit quality: 
$\chi^2=14.6$ for 17 degrees of freedom ($d.o.f.$) (62\% CL). 

The fits have three components: direct $\rho\rt\pipi$ ($\propto |BW_{\rho}|^2$)
and $\omega\rt\pipi$ ($\propto r^2_{\omega}|BW_{\omega}|^2$) contributions
and a $\rho$-$\omega$ interference term.  The contributions from each
component for each fit are listed in Table~\ref{tbl:yields}.

\begin{table}[htb]
\begin{center}
\caption{\label{tbl:yields}
Summary of the results from the $\rho$-$\omega$ interference fit.}
\begin{tabular}{l|c|c|c|c|c}\hline
            & $N_{sig}$    & $r_{\omega}$             & $N_{\rho\rt \pi\pi}$ & $N_{\omega\rt\pi\pi}$ & $N_{\rho-\omega~{\rm interf}}$ \\\hline
$S$-wave    &  $159\pm 15$ & $0.07\pm 0.05$           &    140.9            & $0.6\pm 0.5$         &         17.8           \\
$P$-wave    &  $158\pm 15$ & $0.48^{+0.20}_{-0.14}$   &     93.2            &  $3.6^{+1.5}_{-1.1}$ &         60.0           \\\hline\hline
\end{tabular}
\end{center}
\end{table}

If the low-mass tails of the $\omega\rt\pipi\pi^0$  and $\omega\rt\pipi$ lineshapes
are the same~\cite{pko}, we expect
\begin{equation}
\frac{N(\omega\rt\pi\pi)}{ N_{\rm sig}} 
= \frac{{\mathcal B}(X(3872)\rt \omega\jp)}{{\mathcal B}(X(3872)\rt \pipi\jp)}\times
{\mathcal B}(\omega\rt\pipi),
\end{equation}
where the combined result from Belle~\cite{belle_3pijpsi} and BaBar~\cite{babar_3pijpsi}
(measured using  $\omega\rt\pipi\pi^0$ decays) is 
${\mathcal B}(X(3872)\rt \omega\jp)/{\mathcal B}(X(3872)\rt \pipi\jp)=0.8\pm 0.3$.
Using this, $N_{\rm sig}=159\pm 15$ and ${\mathcal B}(\omega\rt\pipi)=0.0153\pm 0.0013$~\cite{PDG},
we find an expected value $N(\omega\rt\pipi)=2.0\pm 0.8$ events, which is between
the values derived from both the $S$-wave and $P$-wave fits and reasonably consistent with either
case.

\section{Summary}

We report a measurement of the difference in masses of $X(3872)$ mesons produced in
 $B^+\rt K^+\pipi\jp$ and $B^0\rt K^0\pipi\jp$ decays,
\begin{equation}
\Delta M_{X(3872)} = ( -0.69 \pm 0.97~{\rm (stat)} \pm 0.19~{\rm (syst)})~{\rm MeV},
\end{equation}
that is consistent with zero and disagrees with theoretical
predictions based on a diquark-diantiquark model for the
$X(3872)$~\cite{maiani}.
We conclude from this that the same particle is
produced in the two processes and use a fit to the combined
neutral and charged $B$ meson data samples to determine:
\begin{equation}
M_{X(3872)}= (3871.84 \pm 0.27~{\rm (stat)} \pm 0.19~{\rm (syst)})~{\rm MeV}.
\end{equation}
This result agrees with the current PDG world-average value of 
$3871.56\pm 0.22$ MeV~\cite{PDG} and supersedes Belle's earlier
mass measurement~\cite{skchoi_x3872}, which was based on
a 140~fb$^{-1}$ subset of the current data sample.
The width of the $X(3872)$ signal peak is consistent with the experimental
mass resolution and we set a 90\% CL limit on its natural width of
 $\Gamma_{X(3872)} < 1.2$~MeV, improving on the previous limit of 2.3~MeV.

We report a new measurement of the product branching fraction
\begin{eqnarray}
{\mathcal B}(B^+\rt K^+ X(3872))\times {\mathcal B}(X(3872)\rt\pipi\jp)& = &
\nonumber
\\
(8.61 \pm 0.82~{\rm (stat)} \pm 0.52~{\rm (syst)})\times 10^{-6},
\end{eqnarray} 
which supersedes the previous Belle result~\cite{skchoi_x3872}.
The $21.0 \pm 5.7$ signal event yield for $B^0\rt K^0 X(3872)$ translates to
a ratio of branching fractions
\begin{equation}
\frac{{\mathcal B}(B^0\rt K^0 X(3872))}
{{\mathcal B}(B^+\rt K^+ X(3872))} = 0.50 \pm 0.14~{\rm (stat)} \pm 0.04~{\rm (syst)}.
\end{equation}

An examination of the isospin-related $B\rt K\pi^+\pi^0\jp$ channel shows no
evidence for a charged partner to the $X(3872)$ decaying as $X^+\rt\rho^+\jp$
and we determine 90\% CL upper limits on the product branching fractions
${\mathcal B}(B\rt K X^+)\times {\mathcal B}(X^+\rt \rho^+\jp)$ of
$4.2\times 10^{-6}$ and $6.1\times 10^{-6}$ for $K=K^+$ and $K=K^0$, respectively,
for an $X^+$ partner state with mass between 3850~MeV and 3890~MeV.   These
limits are well below expectations for the $X(3872)$ if it is purely a
neutral member of an $I=1$ triplet, in which case decays to the $I_3 =\pm 1$ partners
are favored by a factor of two. 

A comparison of angular correlations among the final state decay products 
finds a good match between data and
MC expectations for $J^{PC}=1^{++}$ with no free parameters (other than the
overall normalization).  The $J^{PC}=2^{-+}$ hypothesis has one complex
free parameter and we found a value for which this hypothesis also matches
the data reasonably well. For this parameter value, the differences between $1^{++}$ and $2^{-+}$ 
expectations are small but non-zero and a three-dimensional analysis
based on the angles that we use could distinguish
between the two cases with the much larger data sets expected at the 
LHCb~\cite{LHCb-PUB-2010-003}, Belle II~\cite{belle2_exp} and
SuperB~\cite{superb} experiments.

Fits to the $M(\pipi)$ mass distribution that only consider contributions from
$\rho\rt\pipi$ decays favor $S$-wave ($J^{P} = 1^{+}$)
over $P$-wave ($J^{P} = 2^{-}$).  However, the addition
of an interfering contribution from isospin-violating
$\omega\rt\pipi$ decays results in acceptable fits
for both the $S$-wave and the $P$-wave hypotheses.  The $P$-wave fit requires a
more substantial
contribution from $\omega\rt\pipi$, but with the current limited statistics
for $X(3872)\rt\pipi\jp$ decays and the poor precision on the ratio  
${\mathcal B}(X(3872)\rt \omega\jp)/{\mathcal B}(X(3872)\rt \pipi\jp)$,
the measured $\omega\rt\pipi$ amplitudes that result from fits to $M(\pipi)$
cannot be used to distinguish between the two possibilities.  This also may be
possible in future experiments.

\section{Acknowledgments}

We thank the KEKB group for the excellent operation of the
accelerator, the KEK cryogenics group for the efficient
operation of the solenoid, and the KEK computer group and
the National Institute of Informatics for valuable computing
and SINET4 network support.  We acknowledge support from
the Ministry of Education, Culture, Sports, Science, and
Technology (MEXT) of Japan, the Japan Society for the 
Promotion of Science (JSPS), and the Tau-Lepton Physics 
Research Center of Nagoya University; 
the Australian Research Council and the Australian 
Department of Industry, Innovation, Science and Research;
the National Natural Science Foundation of China under
contract No.~10575109, 10775142, 10875115 and 10825524; 
the Ministry of Education, Youth and Sports of the Czech 
Republic under contract No.~LA10033 and MSM0021620859;
the Department of Science and Technology of India; 
the BK21 and WCU program of the Ministry Education Science and
Technology, National Research Foundation of Korea,
and NSDC of the Korea Institute of Science and Technology Information;
the Polish Ministry of Science and Higher Education;
the Ministry of Education and Science of the Russian
Federation and the Russian Federal Agency for Atomic Energy;
the Slovenian Research Agency;  the Swiss
National Science Foundation; the National Science Council
and the Ministry of Education of Taiwan; and the U.S.
Department of Energy.
This work is supported by a Grant-in-Aid from MEXT for 
Science Research in a Priority Area (``New Development of 
Flavor Physics''), and from JSPS for Creative Scientific 
Research (``Evolution of Tau-lepton Physics''). S.-K. Choi
acknowledges support from NRF Grant No. KRF-2008-313-C00177
and S.L. Olsen acknowledges support from WCU Grant No. R32-10155.


\begin{thebibliography}{99}
\bibitem{conj} The inclusion of charge-conjugate modes
is always implied.
\bibitem{full_data} This is all of the $\Upsilon(4S)$ data
that was accumulated by the Belle experiment.
\bibitem{psiprime} Here $\psip$ is used to designate the $\psi(3686)$
charmonium resonance.  This is sometimes referred to as $\psi(2S)$.
\bibitem{fks} This is the probability that a $K^0$ decays as a
$K_S$ (0.5) times the world-average branching fraction
${\mathcal B}(K_S\rt\pipi)=0.6920 \pm 0.0005$~\cite{PDG}.
\bibitem{lower_limit}
The fit is restricted to the range $\Gamma_{X(3872)}>0.1$~MeV in
order to avoid the singularity in the BW function at zero.
\end{thebibliography}
\end{document}